\documentclass[twocolumn]{aastex63}

\usepackage{booktabs}
\usepackage{graphicx}
\usepackage{enumitem}

\usepackage{CJKutf8}

\usepackage{algorithm}
\usepackage{ amssymb } 
\usepackage{hyperref}

\usepackage{amsmath}
\usepackage{comment}
\usepackage{tabularx}
\usepackage{multirow}
\definecolor{mycol}{rgb}{0.8, 0.6, 1}

\usepackage{natbib}
\usepackage{verbatim}
\usepackage{mathptmx}
\usepackage{url}
\usepackage{numprint}
\usepackage{ulem}

\newcommand{\msun}{\,{\rm M}_\odot}
\newcommand{\msunh}{\,h^{-1}{\rm M}_\odot}

\newcommand{\s}{\,{\rm s}}
\newcommand{\g}{\,{\rm g}}

\definecolor{mycol}{rgb}{0.2, 0.5, 0.1}
\newcommand{\my}[1]{#1} 
\definecolor{mycol2}{rgb}{0.5, 0.2, 0.1}

\received{October XX, 2023}
\revised{October XX, 2023}
\accepted{October XX, 2023}
\submitjournal{ApJ}

\shorttitle{Merger Tree-based Galaxy Matching: A Comparative Study Across Different Resolutions}
\shortauthors{M. Jung et al.}
\graphicspath{{./}{figures/}}

\begin{document}

\title{Merger Tree-based Galaxy Matching: A Comparative Study Across Different Resolutions}

\author[0000-0002-9144-1383]{Minyong Jung}
\correspondingauthor{wispedia@snu.ac.kr}
\affiliation{Center for Theoretical Physics, Department of Physics and Astronomy, Seoul National University, Seoul 08826, Korea}

\author[0000-0003-4464-1160]{Ji-hoon Kim}
\correspondingauthor{mornkr@snu.ac.kr}
\affiliation{Center for Theoretical Physics, Department of Physics and Astronomy, Seoul National University, Seoul 08826, Korea}
\affiliation{Seoul National University Astronomy Research Center, Seoul 08826, Korea}

\author[0000-0003-4597-6739]{Boon Kiat Oh}
\affiliation{Department of Physics, University of Connecticut, U-3046, Storrs, CT 06269, USA}

\author[0000-0003-4923-8485]{Sungwook E. Hong}
\affiliation{Korea Astronomy and Space Science Institute, 776 Daedeok-daero, Yuseong-gu, Daejeon 34055, Korea}
\affiliation{Astronomy Campus, University of Science and Technology, 776 Daedeok-daero, Yuseong-gu, Daejeon 34055, Korea}

\author[0000-0002-6810-1778]{Jaehyun Lee}
\affiliation{Korea Astronomy and Space Science Institute, 776 Daedeok-daero, Yuseong-gu, Daejeon 34055, Korea}
\affiliation{Korea Institute for Advanced Study, 85 Hoegi-ro, Dongdaemun-gu, Seoul 02455, Korea}

\author[0000-0002-4391-2275]{Juhan Kim}
\affiliation{Center for Advanced Computation, Korea Institute for Advanced Study, 85 Hoegiro, Dongdaemun-gu, Seoul 02455, Korea}







\begin{abstract}
We introduce a novel halo/galaxy matching technique between two cosmological simulations with different resolutions, which utilizes the positions and masses of halos along their subhalo merger tree. With this tool, we conduct a study of resolution biases through the {\it galaxy-by-galaxy} inspection of a pair of simulations that have the same simulation configuration but different mass resolutions, utilizing a suite of {\sc IllustrisTNG} simulations to assess the impact on galaxy properties. We find that, with the subgrid physics model calibrated for TNG100-1, subhalos in TNG100-1 (high resolution) have $\lesssim0.5$ dex higher stellar masses than their counterparts in the TNG100-2 (low-resolution). It is also discovered that the subhalos with $M_{\mathrm{gas}}\sim10^{8.5}\,{\rm M}_\odot$ in TNG100-1 have $\sim0.5$ dex higher gas mass than those in TNG100-2. The mass profiles of the subhalos reveal that the dark matter masses of subhalos in TNG100-2 converge well with those from TNG100-1, except within 4 kpc of the resolution limit. The differences in stellar mass and hot gas mass are most pronounced in the central region. We exploit machine learning to build a correction mapping for the physical quantities of subhalos from low- to high-resolution simulations (TNG300-1 and TNG100-1), which enables us to find an efficient way to compile a high-resolution galaxy catalog even from a low-resolution simulation. Our tools can easily be applied to other large cosmological simulations, testing and mitigating the resolution biases of their numerical codes and subgrid physics models. 


\end{abstract}
\keywords{galaxies:formation --- galaxies:evolution --- galaxies:haloes --- galaxies: statistics --- cosmology: theory --- cosmology:dark matter --- cosmology:large-scale structure of Universe --- methods: numerical --- methods: analytical}



\section{Introduction}

Emerging as powerful tools in modern astrophysics, cosmological simulations enable scientists to study the formation of dark matter (DM) halos, galaxies, and large-scale structures. DM-only (DMO) simulations, paired with semi-analytic models, have achieved significant success in advancing galaxy formation theory \citep[e.g.,][]{1993MNRAS.262..627L}. Furthermore, the increase in computational power has also paved the way for the reproduction of baryonic components --- such as the interstellar medium, stars, and black holes \citep[for a recent review, see][]{2023ARA&A..61..473C}. 
However, even as the resolution of these simulations improves, it remains computationally challenging to fully resolve the intricate physics of many processes of interest. 

Given the complexities of baryon physics and the differences in resolution scales, a purely {\it ab initio} approach is impractical. 
This is where subgrid models become indispensable in cosmological simulations, serving as a mechanism to implement crucial baryonic physics --- star formation and feedback, interstellar medium, massive black hole (MBH) accretion and feedback, and so on --- when the computational resolution is limited. 
The same model, however, can behave differently based on the numerical scheme and resolution employed in the simulation \citep{2015MNRAS.446..521S}. To enhance the reliability of simulations, it is crucial to quantify the impacts of resolution on the subgrid recipes.

\cite{2015MNRAS.446..521S} introduced the terms `weak convergence' and `strong convergence' in the context of numerical simulations' predictive power with respect to observables. 
`Strong convergence' is achieved when the subgrid model yields consistent results across different resolutions. 
However, most observables exhibit only `weak convergence', meaning that the results require re-calibration of the subgrid parameters with changes in resolution to some extent. 
For example, the EAGLE simulation \citep{2015MNRAS.446..521S, 2015MNRAS.450.1937C} exhibits a good weak convergence in the galaxy stellar mass function. 
This means that the convergence of the galaxy stellar mass function is achieved by adjusting the feedback parameters for a different resolution accordingly. However, a strong convergence test --- conducted without calibrating subgrid parameters --- shows that the galaxy stellar mass function at $M_{\star} \sim 10^9 \msun$ in a simulation with 8 times better mass resolution is 0.4 dex higher. \cite{2015MNRAS.446..521S} also tested the convergence of the recalibrated model --- feedback parameters were adjusted to achieve the galaxy stellar mass function  --- in the other properties. They found a good weak convergence in terms of galaxy sizes, black hole mass, and star formation rate for a given galaxy stellar mass in the {\sc EAGLE} simulation. However, for the mass–metallicity relation, the strong convergence test yields significantly better results than the weak convergence test with the recalibrated model. This result suggests that, while weak convergence can be achieved for some properties by adjusting subgrid parameters with resolution, other properties may not converge using these parameters.

Resolution convergence have been studied in large-scale cosmological simulation, using matter power spectrum \citep{2015MNRAS.452.2247V, 2016JCAP...04..047S, 2018MNRAS.477..983S}, gas fraction \citep{2007MNRAS.377...41C, 2013MNRAS.434.2645D, 2017MNRAS.467.1678Q, 2020MNRAS.493.2926L}, internal structures of dark matter halos and stellar kinematics \citep{2019MNRAS.488.3663L, 2020MNRAS.493.2926L,2023arXiv230605753L}, major mergers \citep{2016MNRAS.462.2418S}, and compact galaxies \citep{2020A&A...643L...8C}. 
In the {\sc IllustrisTNG} (TNG hereafter) simulations, \cite{2018MNRAS.473.4077P} investigated the effects of varying numerical resolution while maintaining a fixed subgrid physics model. 
They observed that as the resolution increases, galaxies tend to have higher stellar masses. 
They also found that, for a given stellar mass, the increased resolution results in higher black hole mass and metallicity. In their highest resolution run, TNG50, a convergence test was conducted on morphological properties such as galaxy sizes, disk heights, and kinematics \citep{2019MNRAS.490.3196P}. 
Comparisons of various other properties, such as stellar morphologies and star-forming activities \citep{2021MNRAS.501.4359Z} and the radial distribution of satellite galaxies \citep{2022MNRAS.514.4676R} with simulations of different resolutions within the TNG suite have also been performed.

However, most of these resolution studies have been done through several key scaling-relation comparisons, not galaxy-by-galaxy comparisons. This is primarily due to the difficulty of matching halos and galaxies in simulations with different resolutions. It is therefore notable that in comparison studies between hydrodynamic simulations and their DMO counterparts --- where the number of dark matter particles is identical in both simulations --- researchers have been able to directly investigate the role of baryonic physics by matching galaxies across the two types of simulations \citep[e.g.,][]{2013MNRAS.431.1366S,2015MNRAS.451.1247S}.
\cite{2022MNRAS.509.5046L} adopted a similar matching technique, and applied a machine learning (ML) method to correct the differences induced by baryonic physics. 
These comparison studies were possible because DMO simulations, in most cases, start from the same initial conditions and with the same DM particle IDs as used in the hydro simulations. If the DM particles in both the hydrodynamic and DMO simulations share the same particle numbering scheme, matching pairs of halos can be readily identified by locating the halo containing particles with same particle IDs \citep[e.g., as demonstrated by][]{2015MNRAS.451.1247S}. This indicates that they originated from the same patch region in the initial conditions.

\begin{table*}[ht]
\centering
\vspace{1mm}
\begin{tabular}{@{}p{1.8cm}p{1.8cm}p{1.2cm}p{1.8cm}p{1.8cm}p{1.8cm}p{1.9cm}p{1.8cm}@{}}
\toprule
Name    & $L_{\rm box}$ [cMpc]& $N_{\rm DM}$ & $m_{\rm DM}$ [$\msun$] & $m_{\rm gas}$ [$\msun$] & $\epsilon_{\rm DM,\,star}^{z=0}$ [kpc] & $\epsilon_{\rm gas,\,min}$ [ckpc] & $\bar{r}_{\rm cell}$ [pc] \\ %
\midrule 
TNG50-1  & 51.7   & $2160^3$ & $4.5 \times 10^5$   & $8.5\times 10^4$ & 0.29 &$0.074$ &  5.8  \\
TNG50-2  & 51.7   & $1080^3$ & $3.6 \times 10^6$   & $6.8\times 10^5$ & 0.58 &$0.148$ &  12.9 \\  
TNG100-1 & 106.5  & $1820^3$ & $7.5 \times 10^6$ & $1.4 \times 10^6$ & 0.74 &$0.185$ &  15.8 \\
TNG100-2 & 106.5  & $910^3$ & $6.0 \times 10^7$ & $1.1 \times 10^7$  & 1.48 &$0.369$ &  31.2 \\
TNG300-1 & 302.6  & $2500^3$ & $6.0 \times 10^7$ & $1.1 \times 10^7$  & 1.48 &$0.369$ &  31.2 \\%
\bottomrule
\end{tabular}
\vspace{1mm}
\caption{The key  parameters for the TNG simulation suite used in our study.  {\it From left to right:} the box length (in a comoving unit), number of DM particles, DM particle mass, target baryon mass, Plummer equivalent gravitational softening length of the DM and stellar particles, minimum value of the adaptive gravitational softening length for gas (in a comoving unit), and median gas cell radius. Note that the gas cell mass has continuous values centered around the target baryon mass, $m_{\rm gas}$, within a factor of two \citep{2018MNRAS.473.4077P}. See \cite{2019MNRAS.490.3196P}, \cite{2019MNRAS.490.3234N}, or the \href{https://www.tng-project.org/data/docs/background/}{TNG Collaboration website} for a detailed list of the parameters.}
\label{tab:tng}
\vspace{2mm}
\end{table*}

Applying this matching technique directly to pair halos/galaxies in high- and low-resolution simulations is challenging because the numbers of DM/gas particles are different in the two simulations. 
In such cases, one possible solution is to trace each particle back to its initial conditions and identify the nearest counterpart particle in the other simulation \citep[e.g., ][]{2023arXiv230605753L}. Instead, we develop a physically-motivated matching algorithm that utilizes the positions and masses of halos along the merger tree in the matching process. Using this algorithm, we for the first time perform a {\it galaxy-by-galaxy} resolution study with our matching halo/galaxy catalogs, and investigate how the galaxy properties differ with varying resolutions in the TNG simulation suite. 
Moreover, inspired by ML techniques used to paint baryonic properties onto DM halos in DMO simulations \citep[e.g.,][]{2019MNRAS.489.3565J,2022MNRAS.509.5046L,2022ApJ...941....7J}, we seek to adjust the physical properties of halos found in low-resolution simulations.
As a first test, we implement an ML model for TNG300-1 subhalos which ``corrects'' their properties to match those in a higher resolution simulation, TNG100-1. 

The remainder of this paper is organized as follows. In Section \ref{sec:method}, we describe the  simulation suite used, the subhalo matching technique, and a brief illustration of the ML model employed. 
In Section \ref{sec:result} we demonstrate that, while DM halo properties exhibit strong resolution convergence, the baryonic properties change with resolution (Section \ref{sec:result_halo}). 
We further investigate the differences in the radial mass profiles of the subhalos (Section \ref{sec:result_rc}). 
We then apply an ML method to ``correct'' the resolution biases in stellar mass, gas mass, and metallicity (Section \ref{sec:ml_result}). 
Lastly, in Section \ref{sec:discussion}, we discuss the validity of our matching catalogs and analyze which halo features are more influential in our ML model.

\section{Methodology}\label{sec:method}

\subsection{{\sc IllustrisTNG} Simulation Suite}\label{sec:tng}

{\sc IllustrisTNG} (TNG in short) is a series of large cosmological simulations using the moving-mesh code {\sc Arepo} \citep{2010MNRAS.401..791S}. 
TNG implemented various sophisticated subgrid physics model including magnetohydrodynamics \citep{2011MNRAS.418.1392P,2013MNRAS.432..176P}, black hole accretion and feedback \citep{2017MNRAS.465.3291W}, galactic winds and stellar evolution \citep{2013MNRAS.436.3031V,2018MNRAS.473.4077P}, and metal advection \citep{2018MNRAS.477.1206N, 2018MNRAS.473.4077P}.  
In this paper, we primarily employ four simulations in their suite: TNG100-1, TNG100-2, TNG50-1, and TNG50-2. 
TNG100-1 served as the fiducial simulation for which the subgrid physics model parameters have been calibrated, with a box size of (106.5 cMpc)$^3$ \citep{2019ComAC...6....2N}.\footnote{For the calibration, the cosmic star formation rate density as a function of time, the galaxy stellar mass function at $z = 0$, and the stellar-to-halo-mass relation at $z = 0$ are used \citep{2018MNRAS.473.4077P}.} 
The physics model and parameters used in TNG100-1 were implemented in all TNG simulations. 
TNG50-1 was the highest resolution run in the TNG suite, with a smaller box size of (51.7 cMpc)$^3$. 
TNG100-2 (TNG50-2) was a lower-resolution counterpart of TNG100-1 (TNG50-1). 
With the same initial condition, each particle mass in TNG100-2 (TNG50-2) is 8 times higher than the flagship run, TNG100-1 (TNG50-1). 
The mass and spatial resolutions of these simulations are listed in Table \ref{tab:tng}, including TNG300-1 which is used for the analysis in Section \ref{sec:ml_result}. 
To construct the matching algorithm in Section \ref{sec:matching}, we employ TNG100-1-Dark and TNG50-1-Dark, the DMO counterpart simulation of TNG100-1 and TNG50-1, respectively. 
The pair of simulations have an identical number of DM particles, but lack the baryon components. 
Therefore, the DM particle mass in TNG100-1-Dark and TNG50-1-Dark is $\Omega_{\rm matter}/\Omega_{\rm DM}$ times higher than that in the hydrodynamic simulations. 

TNG implemented different softening lengths by mass resolution, following $\epsilon_{\rm DM, \,stars} = L_{\rm box}/N_{\rm DM}^{1/3}/40$ for DM and stellar particles, where $L_{\rm box}$ is the simulation box length and $N_{\rm DM}$ is the total number of DM particles \citep{2018MNRAS.473.4077P}. 
The adaptive softening length of gas followed $\epsilon_{\rm gas} = 2.5\,r_{\rm cell}$, where $r_{\rm cell}$ is the effective radius of a gas cell \citep{2018MNRAS.473.4077P}. 
The black hole kernel-weighted neighbor number was also scaled with the resolution as $n_{\rm ngb}  \propto m_{\rm baryon}^{-1/3}$ \citep{2017MNRAS.465.3291W}. 
In addition, due to numerical reasons, the physics model for TNG50 was slightly modified --- specifically, the dependence of the star formation timescale on gas density \citep[see Section 2.2 in][]{2019MNRAS.490.3234N}. 
 Except for these listed above, all the physics models are identical across all TNG simulations, regardless of the resolution. 
Most notably, all TNG simulations adopt the same star formation density threshold, $n_{\rm H} \simeq 0.1\, {\rm cm}^{-3}$, above which stochastic star formation occurs \citep{2018MNRAS.473.4077P}. 

\begin{figure*}
    \centering
    \vspace{0mm}    
    \includegraphics[width=\linewidth]{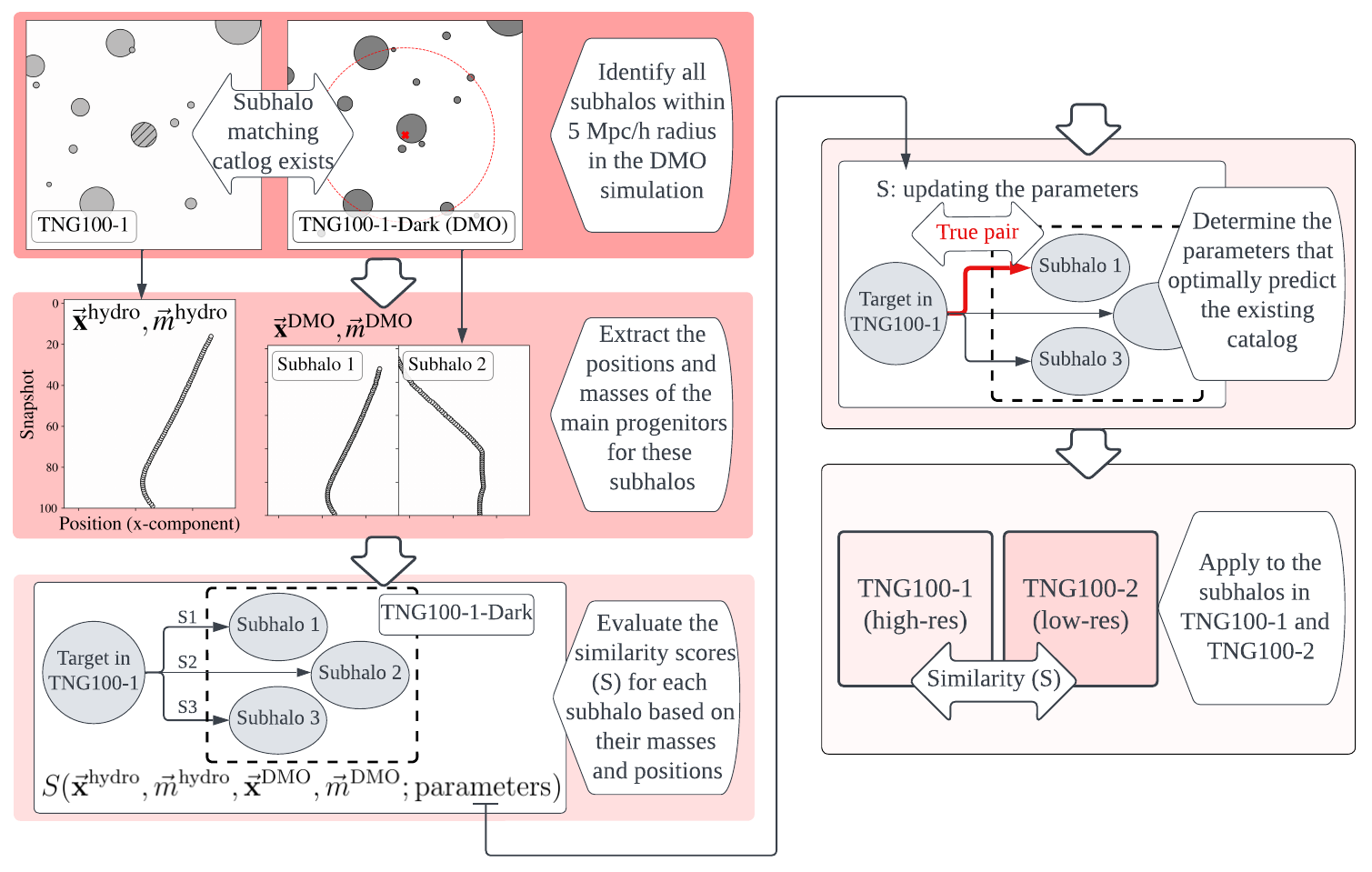}    
    \vspace{-2mm}        
    \caption{Schematic diagram of the subhalo matching method between TNG100-1 (high-resolution) and TNG100-2 (low-resolution) using machine learning (ML). 
    We start from subhalos in TNG100-1 and their DMO counterparts in TNG100-1-Dark, where the subhalo matching catalog already exists ({\it first panel from top left}). 
    We extract positions and masses of each subhalo along the main branch of their merger tree, and evaluate the similarity scores for all possible pairs between subhalos in TNG100-1 and TNG100-1-Dark ({\it second and third panel}). 
    The matching function is governed by several parameters as described in Eqs.(\ref{equation_train}) and (\ref{equation_train2}). 
    After we tune the parameters that optimally predict the existing catalog ({\it fourth panel}), the matching function with the optimal set of parameters is used to construct a matching catalog between the subhalos in TNG100-1 and TNG100-2 ({\it fifth panel}). 
    See Section \ref{sec:matching} and Appendix \ref{sec:matching_appendix} for a detailed description of the pipeline.}
   \vspace{4mm}    
\end{figure*}

DM halo catalogs distributed by the TNG Collaboration were compiled using halos identified by the Friends-of-Friends algorithm \citep[FoF;][]{1985ApJ...292..371D}. 
The subhalos, which are self-bounded substructures like the central and satellite galaxies within the halos, were further identified using the {\sc SubFind} algorithm \citep{2001MNRAS.328..726S, 2009MNRAS.399..497D}. In this work, the gas mass and stellar mass of the subhalos represent the total mass of components bound to the subhalos, determined by the {\sc SubFind} algorithm. 
From the snapshot 0 ($z=20.05$) to snapshot 99 ($z=0$), subhalos across cosmic times were linked together in a merger tree using two different algorithms: {\sc LhaloTree} \citep{2005Natur.435..629S} and {\sc SubLink} \citep{2015MNRAS.449...49R}. 
For a detailed comparison of the similarities and differences between the two algorithms, refer to \cite{2015A&C....13...12N}. 
Unless specified, we use the {\sc SubLink} merger trees for our analyses.

Since the DM particle IDs in the hydrodynamic simulations and their DMO counterparts are numbered in exactly the same way, the TNG Collaboration provides subhalo matching catalogs between the hydrodynamic and DMO runs \citep{2015A&C....13...12N}. 
For each subhalo in the hydrodynamic simulation, the subhalo with the largest number of the shared DM particles in their {\sc LhaloTree} merger trees in the DMO run was chosen to be the best-matched candidate. 
After repeating the process, but this time finding a best-matched candidate in a hydrodynamic run for a subhalo in the DMO run, only the matching pairs that were bijectively matched (i.e., both directions yield identical matching results) were saved.

\subsection{Matching Subhalos Between Simulations}\label{sec:matching}

\begin{figure*}
    \centering
    \includegraphics[width=0.96\linewidth]{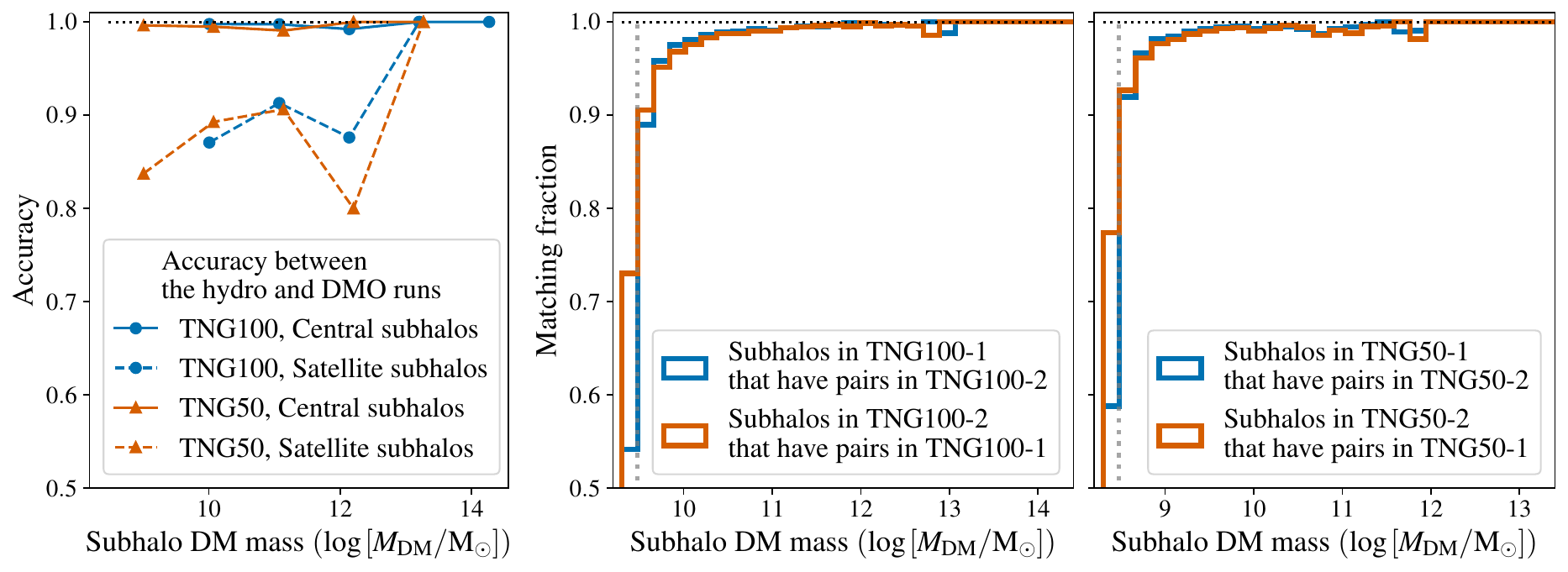}    
    \caption{({\it left}) 
    The accuracy of the matching method compared to the existing subhalo matching catalog. Matching accuracies for subhalo matching between TNG100-1 and TNG100-Dark are represented by {\it blue circles}, and those between TNG50-1 and TNG50-Dark by {\it orange triangles}. Central subhalos (the primary subhalo in its FoF group) and satellite subhalos are differentiated by {\it solid} and {\it dashed lines}, respectively. The matching accuracies of central subhalos in both simulation pairs remain $\sim0.995$ across all subhalo mass ranges, while the matching accuracies for satellite subhalos are relatively lower. Note that at the high mass end, with $M_{\rm DM} > 10^{14} \msun$, only central subhalos in TNG100-1 exist in the test set.     
    ({\it middle} and {\it right}) The fraction of subhalos in the high-resolution simulation for which their counterparts in the low-resolution simulation are identified by our subhalo matching method (and vice versa). 
    The {\it gray vertical lines} indicate $3 \times 10^9 \msun$ and $3 \times 10^8 \msun$ in the {\it middle} (for TNG100) and {\it right panel} (for TNG50), respectively. 
    For over 98\% of subhalos with mass $M_{\rm halo} > 10^{10} \msun$ in TNG100-1 (TNG100-2), their matching pairs are identified in TNG100-2 (TNG100-1). See Section \ref{sec:matching} for more details.}
    \label{fig:matching_hist_lowres}
\vspace{1mm}    
\end{figure*}

To construct and test the matching algorithm, we start with TNG100-1 and its counterpart DMO run, TNG100-1-Dark, where matching catalogs between the two simulations are provided by the TNG Collaboration (Section \ref{sec:tng}). 
During this training phase, our aim is to determine the optimal parameters for the galaxy/subhalo matching model between two simulations.\footnote{The basic assumption here is that the optimal ML model for matching subhalos between high- and low-resolution simulations is very close to that between hydrodynamic and DMO simulation. However, the optimal set of parameters could be somewhat different between the two pairs of simulations.  Therefore, in order to acquire a solid performance, we impose a very strict criteria as we build our training set, selecting only the subhalo pairs that are bijectively matched between two simulations (Section \ref{sec:tng}), and those with prediction probabilities higher than 0.7 (Appendix \ref{sec:matching_appendix}).}

We first evaluate the ``similarity'' between a subhalo in TNG100-1 (denoted as {\it ``hydro''}) and a subhalo in TNG100-1-Dark (denoted as {\it ``DMO''}). This evaluation begins with acquiring the positions and masses of all the halos at $z=0$ along the main branches of their merger trees, $\bold{x}_{I}^{\rm hydro}, \bold{x}_{I}^{\rm DMO}, m_{I}^{\rm hydro},$ and $m_{I}^{\rm DMO}$, where $I$ indicates a snapshot number (ranging from $N_{\rm start}$ to 99, with $N_{\rm start}$ indicating the snapshot of the last leaf on main branch). 
Our goal is then to determine whether the two subhalos have similar positions and DM masses at all times along the main branches of their respective merger trees. 
Specifically, we employ the following scoring scheme to evaluate the similarity $S$ between the two halos:
\begin{equation}\label{equation_train}
S = - \sum_{I=N_{\rm start}}^{99} w_If(I) + b,    
\end{equation}
where
\begin{multline}\label{equation_train2}
f(I)\equiv \frac{ |\bold{x}_{I}^{\rm hydro} - \bold{x}_{I}^{\rm DMO}|}{100\,h^{-1}{\rm kpc}} + w_{\rm mass}\left|\log{\left(\frac{m_{I}^{\rm hydro}}{m_{I}^{\rm DMO}}\right)}\right|^{\gamma}. 
\end{multline}
Here, $\gamma$, $w_{\rm mass}$, $w_{I}$, and $b$ are parameters that we have trained using TNG100-1 and TNG100-1-Dark pairs, choosing parameters that optimally predict the ``true" pair in the existing catalog provided by the TNG Collaboration.\footnote{The parameter $b$, known as the bias term, does not have a direct physical meaning. Instead, it is utilized when computing the probability with a {\it softmax} function, that a given subhalo is a true counterpart.} 
We calculate the distance, $|\bold{x}_{I}^{\rm hydro} - \bold{x}_{I}^{\rm DMO}|$, and mass ratio, $|m_{I}^{\rm hydro}/m_{I}^{\rm DMO}|$, in each snapshot $I$, for all progenitors along the main branch of the merger tree. 
If a subhalo in the DMO simulation has no progenitors in a certain snapshot (e.g., when the subhalo formed at a later epoch), we impose a penalty value of $f(I) = p$ to that pair.
The value of $p$ is also determined during the training phase. 
The score $S$ close to zero means that the two subhalos in two simulations formed at similar epoch, and have almost identical growth histories in terms of their DM masses and positions. 

In the left panel of Figure \ref{fig:matching_hist_lowres}, we present the accuracy of matched pairs, defined as the fraction of predictions that are identical to those from the existing catalog, tested between the hydro and DMO simulations. For verification purposes, we allocate 60\% of the simulation volume to the training set and 20\% each to the validation and test sets. The results from the test set indicate that, for the pairs predicted by our matching algorithm, 99.6\% of the central subhalos are matched identically with the catalog provided by the TNG Collaboration. However, the satellite subhalos exhibit relatively lower accuracy, with 87.5\% of subhalo pairs in TNG100 and 84.3\% in TNG50 being correctly matched. It is noteworthy that most of the incorrect matches are attributed to subhalos lacking counterparts in the existing matching catalogs. Our model shows a limitation in eliminating those subhalos from the catalogs, occasionally pairing with subhalos considered to have a similar growth history. A similar, possibly lower level of accuracy is expected when the model is applied to actual datasets, comparing the high- and low-resolution simulations. In Section \ref{sec:verifi}, we further discuss that most of those pairs we match originate from a similar dark matter patch in the initial condition. However, for some pairs, the extent of the similarity --- measured by the fraction of shared DM particles --- is insufficient compared to those in the existing catalog by the TNG Collaboration.

\begin{table*}[]
\vspace{1mm}
\centering
\begin{tabular}{@{}ll@{}}
\toprule
\textbf{Feature name} & \textbf{Description} \\ \midrule
\texttt{SubhaloMassType1}$^{(a)(b)}$ & Total DM mass of the subhalo\\  
\texttt{SubhaloMassType4}$^{(a)}$ & Total stellar mass of the subhalo\\
\texttt{SubhaloMassType0}$^{(a)}$ & Total gas mass of the subhalo \\
\texttt{SubhaloBHMass}$^{(a)}$ & Total black hole mass of the subhalo \\
\texttt{SubhaloVmax}$^{(a)(b)}$ &  Maximum value of the spherically-averaged rotation velocity \\
\texttt{SubhaloVelDisp}$^{(a)(b)}$ &  One-dimensional velocity dispersion of all the member particles/cells\\
\texttt{SubhaloSpin}$^{(a)}$ &  Magnitude of the subhalo total spin\\
\texttt{VelSpinAngle} & Cosine similarity between peculiar velocity and spin \\
\texttt{SubhaloVel}$^{(a)(b)}$ & Magnitude of the subhalo peculiar velocity \\
\texttt{SubhaloStarMetallicity}$^{(a)}$ & Mass-weighted average metallicity of the star particles\\
\texttt{SubhaloGasMetallicity}$^{(a)}$ & Mass-weighted average metallicity of the gas cells within twice the stellar half mass radius\\
\texttt{SubhaloGasMetallicityMaxRad}$^{(a)}$ & Mass-weighted average metallicity of the gas cells within the maximum circular velocity radius\\
\texttt{SubhaloHalfmassRadType1}$^{(a)(b)}$ & Radius containing half of the DM mass of the subhalo\\
\texttt{SubhaloHalfmassRadType4}$^{(a)}$ &  Radius containing half of the stellar mass of the subhalo \\
\texttt{SubhaloHalfmassRadType0}$^{(a)}$ &  Radius containing half of the gas mass of the subhalo  \\
\texttt{SubhaloSFR}$^{(a)}$ & Sum of the star formation rates (SFRs) of all the gas cells in the subhalo  \\
\texttt{GroupMassType1}$^{(a)(b)(c)}$ & DM mass of the host halo containing the target subhalo \\
\texttt{GroupMassType4}$^{(a)(c)}$ &  Stellar mass of the host halo containing the target subhalo \\
\texttt{GroupMassType0}$^{(a)(c)}$ &  Gas mass of the host halo containing the target subhalo \\
\texttt{Group\_R\_TopHat200}$^{(a)(b)(c)}$ & Virial radius of the host halo containing the target subhalo  \\
\texttt{GroupGasMetallicity}$^{(a)(c)}$ & Mass-weighted average metallicity of the host halo containing the target subhalo   \\
\texttt{HostDistance}$^{(b)(c)}$ &  Distance between the subhalo and center of the host halo\\
\texttt{MainProgenitorMassFraction}$^{(b)(c)}$ & Mass fraction of the main progenitor among all progenitors \\
\bottomrule
\end{tabular}
\caption{The list of features used in our ML model to correct the resolution biases of low-resolution subhalos to match their high-resolution counterparts. 
We use these features along 100 snapshots of the main branch of each subhalo's merger tree. 
The features marked with a superscript {\it ``(a)''} follow the same definition described in the TNG webpage, \href{https://www.tng-project.org/data/docs/specifications/}{https://www.tng-project.org/data/docs/specifications/}. For the ablation study in Section \ref{sec:feature_importance}, we classify some feature into the following types: the features marked with {\it ``(b)''} can also be obtained in the DMO simulations, and the features marked with {\it ``(c)''} contain local environmental information.}
\label{tab:features}
\vspace{0mm}
\end{table*}

During the application phase, we utilize the trained machine to match subhalo pairs between high-resolution (TNG100-1) and low-resolution (TNG100-2) simulations. 
By pairing the two subhalos with the highest similarity, we construct the matching catalog between the simulations with two different resolutions. 
We apply the same matching process for TNG50-1 and TNG50-2. 
The fraction of subhalos that have a counterpart in the other simulation is shown in Figure \ref{fig:matching_hist_lowres}. 
We are able to identify matching subhalos for over 98\% of those with $M_{\rm DM} > 10^{10} \msun$ in TNG100-1 or TNG100-2 (or $M_{\rm DM} > 10^{9} \msun$ in TNG50-1 or TNG50-2). 
In contrast, the fraction drops dramatically for subhalos close to the halo mass threshold of $2\times 10^9 \msun$ or 33 DM particles for TNG100 (or, $2\times 10^8 \msun$ or 55 DM particles for TNG50). 
This drop occurs mainly due to the cases where one of the subhalo pairs falls below the mass threshold. 
Therefore, for most of the subsequent analyses we employ the matched pairs in which DM masses in both of the simulations exceed $3\times 10^9\msun$ or 50 DM particles for TNG100 (or, $3\times 10^8\msun$ or 82 DM particles for TNG50; the gray vertical lines in Figure \ref{fig:matching_hist_lowres}). We refer the interested readers to Appendix \ref{sec:matching_appendix} for the detailed explanation of the matching process.

\subsection{Machine Learning Model To Correct Low-resolution Subhalos}\label{sec:correcting}

Because the properties derived from lower-resolution galaxies are biased, we compensate for the resolution difference with a model trained on the high-resolution simulations. We employ {\sc LightGBM} \citep{NIPS2017_6449f44a}, a modified version of the gradient tree boosting algorithm, and construct three ML models --- each predicting stellar mass, gas metallicity, and gas mass. 
In the algorithm, rather than training a single decision tree, an ensemble of decision trees is iteratively trained. 
Each subsequent tree emphasizes correcting the mistakes of its predecessor by assigning greater weight to examples that the previous tree found challenging to predict. 
However, unlike traditional gradient boosting algorithms which build trees level (depth)-wise, {\sc LightGBM} employs a leaf-wise growth strategy, splitting the leaf with the maximum loss reduction. 
For an in-depth discussion of {\sc LightGBM} and the tree boosting algorithms, see \cite{NIPS2017_6449f44a}.

For our input data, we utilize 23 features for each snapshot, encompassing both the subhalo properties and their host halo properties. 
The features used are detailed in Table \ref{tab:features}. 
It is worth to note that, inspired by the previous studies that either preprocessed the temporal history of a target halo into tabular data \citep{2019MNRAS.489.3565J}, or those that used the merger trees themselves as inputs \citep{2022MNRAS.513.5423M, 2022ApJ...941....7J}, we extract all 23 features for {\it all} the main progenitors of the target subhalo. 
That is, we use a total of 2300 features (23 features $\times$ 100 snapshots) as inputs for a single subhalo. 
For the machine training we use the matching subhalo catalogs from TNG100-1 and TNG100-2 (Section \ref{sec:matching}). 
We divide the subhalo pairs into 80\% for a training set and 20\% for a test set. 
A 3-fold cross-validation method is used for the hyperparameter tuning during the training.

\begin{figure*}
    \vspace{1mm}    
    \centering
    \includegraphics[width=1\linewidth]{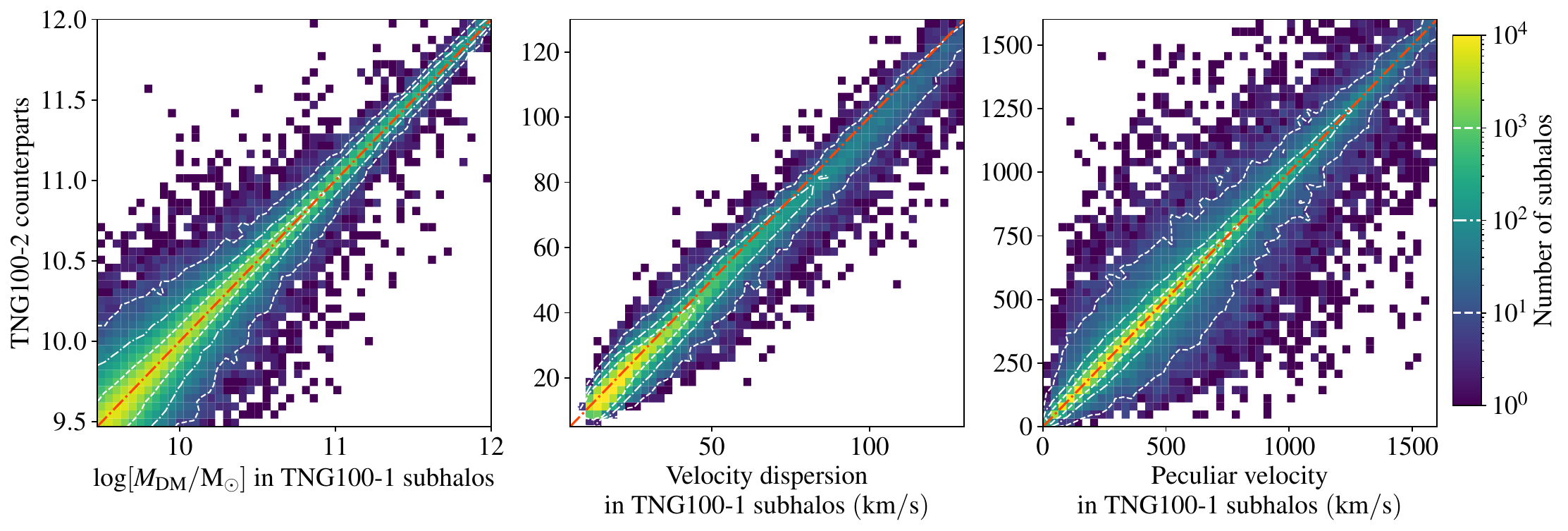}
    \caption{Subhalo properties at $z=0$ that are primarily influenced by DM and gravity, depicted as a two-dimensional histogram. 
    The $x$-axis represents the values of subhalos in TNG100-1 (high-resolution run), while the $y$-axis represents the corresponding values of their matched subhalos in TNG100-2 (low-resolution run). 
    {\it From left to right}, the dark matter mass, velocity dispersion, and peculiar velocity of the subhalos are presented.  
    The color bar denotes the number of subhalos in each bin. 
    The {\it red dot-dashed line} in each panel indicates a perfect correspondence between two simulations. 
    All three properties exhibit good convergence between the simulations of two different resolutions, although the subhalos in the high-resolution run (TNG100-1) have a greater velocity dispersion at the high-mass end.
    See Section \ref{sec:result_halo} for more details.}
    \label{fig:halo_feat}
\end{figure*}

\begin{figure*}
    \vspace{1mm}    
    \centering
    \includegraphics[width=1\linewidth]{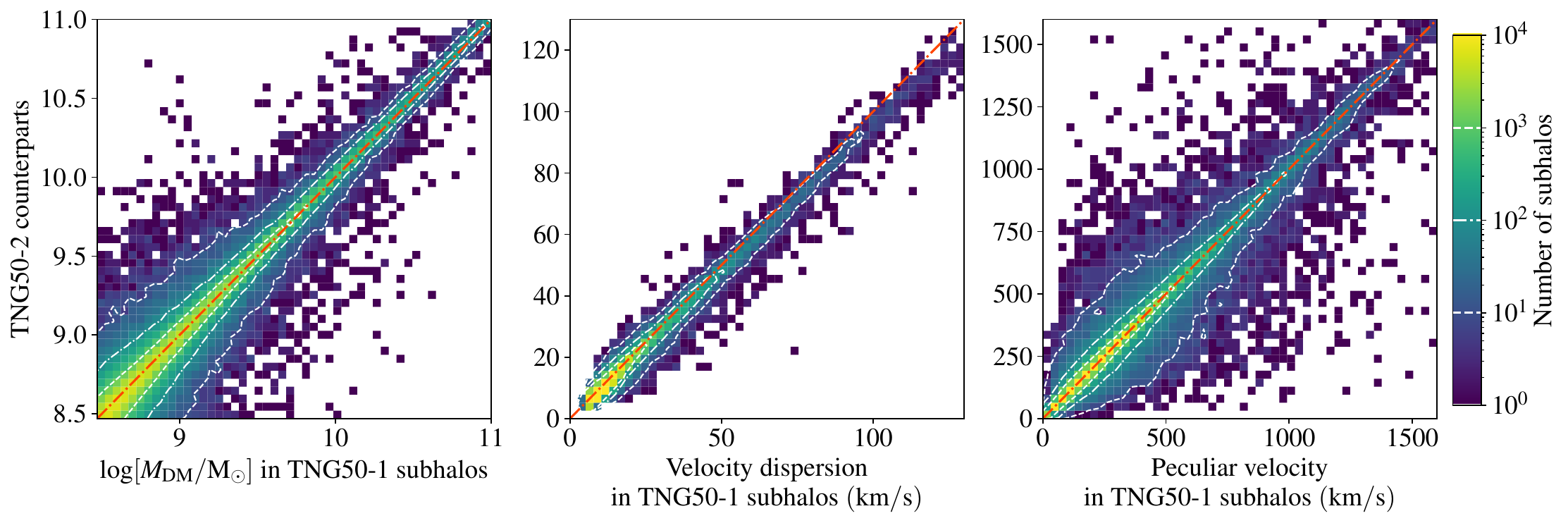}    
    \caption{Same as Figure \ref{fig:halo_feat}, but for TNG50-1 (high-resolution run) and TNG50-2 (low-resolution run). Again, the three properties show good convergence with the varying resolutions.}
    \label{fig:halo_feat-50}
    \vspace{3mm}        
\end{figure*}

As we will discuss in Section \ref{sec:result_halo}, a significant number of subhalo properties at the lower-mass end of the subhalo sample have null values, due to the resolution limit of the simulation. 
To properly handle our data with frequent zero-valued properties (so-called ``zero-inflated'' data), we adopt a two-step approach. 
We develop a binary classification model that predicts whether a property (e.g., the gas mass of the subhalo) is zero or not. 
For this model, we use the AUC-ROC score (Area Under the Curve -- Receiver Operating Characteristic) as a metric. 
Then, for the subset that the initial model identifies as possessing a physical component, we apply a regression model to predict the actual value of the property, using the mean squared error (MSE) as a performance metric.

\section{Results}\label{sec:result}

\subsection{Galaxy-by-galaxy Resolution Study On Key Halo Properties}\label{sec:result_halo}

\begin{figure*}
    \vspace{1mm}    
    \centering
    \includegraphics[width=0.88\linewidth]{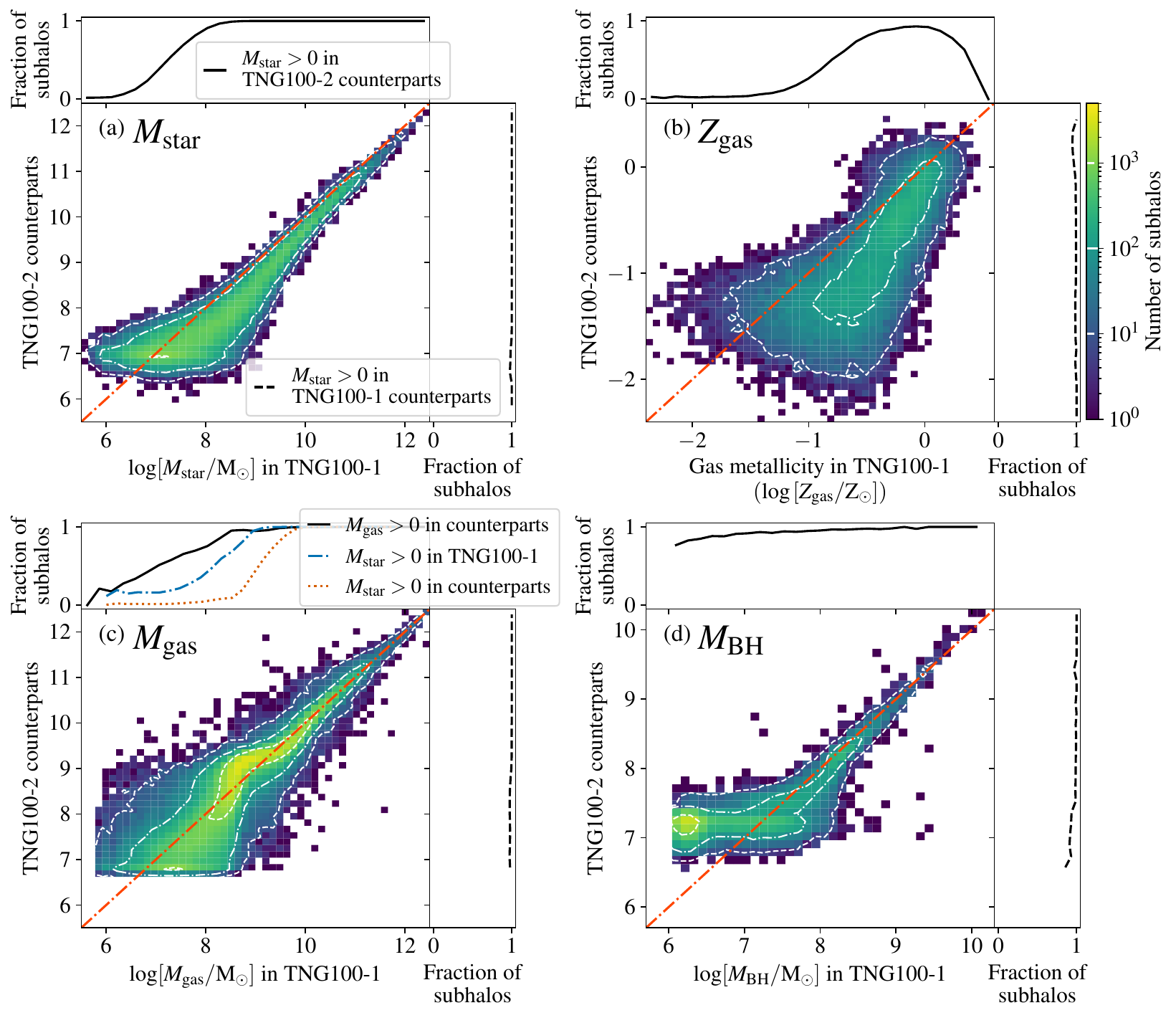}    
    \caption{Baryonic properties of the matching subhalos at $z=0$ in TNG100-1 (high-resolution run) and TNG100-2 (low-resolution run), illustrated as a two-dimensional histogram: {\it (a)} the stellar mass, {\it (b)} the gas metallicity, {\it (c)} the gas mass, and {\it (d)} the black hole mass of the subhalos. 
    The {\it red dot-dashed line} in each panel indicates a perfect convergence with respect to numerical resolution. 
    Due to the higher star formation rates in TNG100-1 with higher resolution, both the stellar mass and the gas metallicity in TNG100-1 are higher than those in TNG100-2. 
    Note that some subhalos are not included in these histograms since they lack either a stellar, gas, or black hole component in one of the two simulations. 
    The {\it black solid/dashed lines} in the {\it marginal plots} at the top and right of each panel indicate the fraction of subhalos for which a given component is present in the counterpart simulation. 
    The {\it blue dot-dashed line} in the top marginal plot of the panel {\it (c)} depicts the fraction of subhalos that have any stellar particle in TNG100-1. 
    Similarly, the {\it orange dotted line} shows the fraction of subhalos that have a positive stellar mass in their TNG100-2 counterparts for a given $M_{\rm gas}$ in TNG100-1. 
    Here one can observe that, in the subhalos of $M_{\rm gas} \sim 10^{8.5}\msun$, star formation occurs only in TNG100-1 but not in TNG100-2.  
    This entails a lack of stellar feedback in TNG100-2 that would have otherwise expelled the gas. 
    Consequently, the gas mass in TNG100-2 is higher than TNG100-1 at $M_{\rm gas} \sim 10^{8.5}\msun$. 
    See Section \ref{sec:result_halo} for more in-depth discussion.}
    \label{fig:mass_feat}
    \vspace{3mm}        
\end{figure*}

\begin{figure*}
    \vspace{1mm}    
    \centering
    \includegraphics[width=0.88\linewidth]{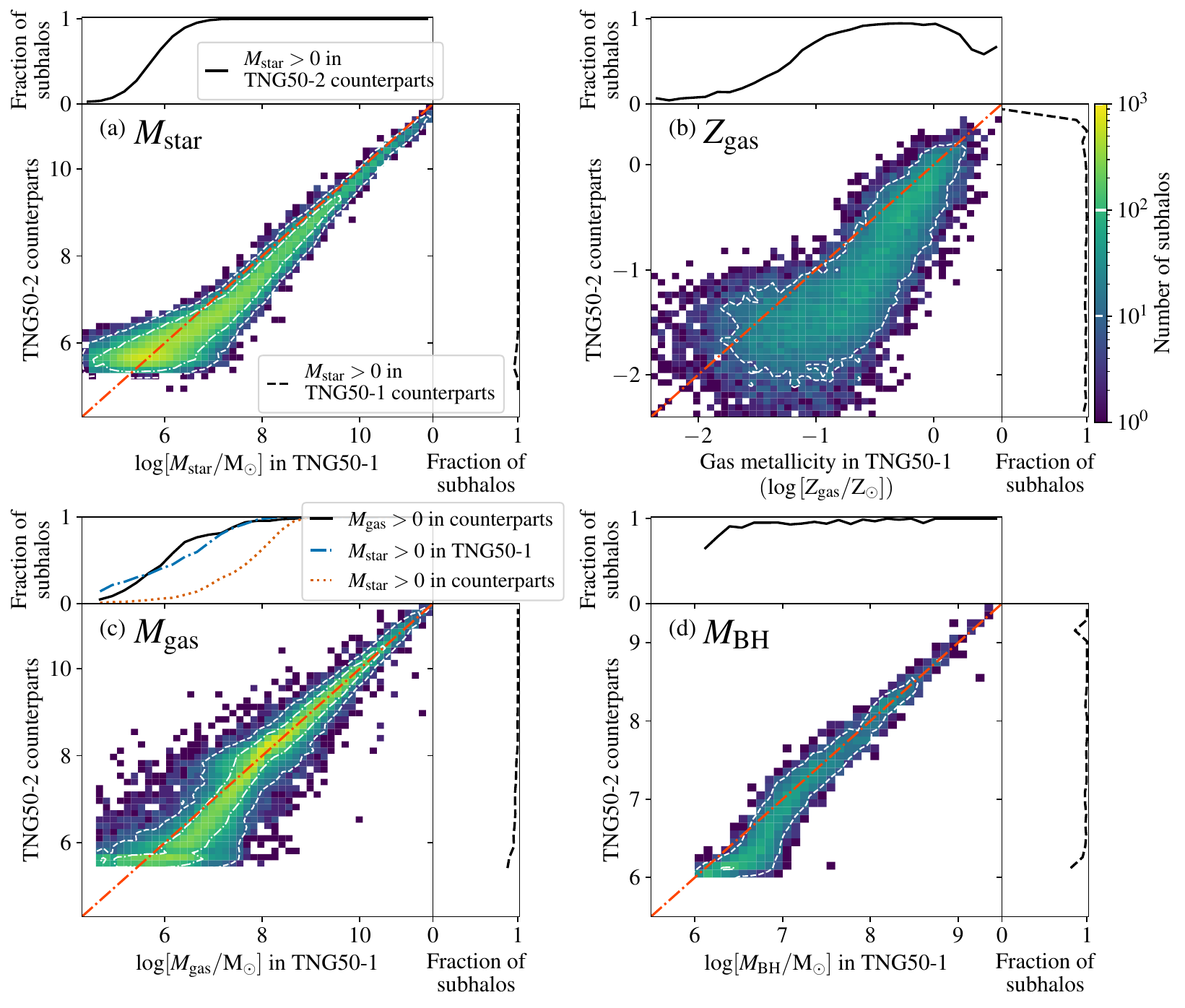}    
    \caption{Same plot as Figure \ref{fig:mass_feat}, but for TNG50-1 (high-resolution run) and TNG50-2 (low-resolution run). While we observe similar trends with the change in resolution as those presented in Figure \ref{fig:mass_feat}, the differences in gas masses are less pronounced. See Section \ref{sec:result_halo} for more details. }
    \label{fig:mass_feat-50}
    \vspace{3mm} 
\end{figure*}

In this section, we analyze the differences in subhalo properties between the high-resolution simulation (TNG100-1) and its low-resolution counterpart (TNG100-2; see Section \ref{sec:matching}). 
In Figure \ref{fig:halo_feat}, we show how three key halo properties that are primarily influenced by DM and gravity --- DM mass, velocity dispersion, and peculiar velocity --- differ between TNG100-1 and TNG100-2. 
These properties show very similar values even when the resolution changes, although the subhalos' velocity dispersions in TNG100-1 are $\sim 4$\% higher than in TNG100-2, especially at the high-mass end. 
This agreement suggests that these subhalo properties  exhibit robust convergence to changes in DM mass resolution.\footnote{We suspect that the baryonic influence is the cause of the slight disagreement in velocity dispersions, as they are computed using all matter components within each  subhalo.} 
We conduct the same analysis for the TNG50-1 and TNG50-2 subhalo pairs, as shown in Figure \ref{fig:halo_feat-50}. 
We find similar agreements with the varying resolution.

Although we observe a robust ``resolution-wide” convergence for those three subhalo properties, there are noticeable systematic discrepancies in baryonic characteristics between different resolutions. 
In Figure \ref{fig:mass_feat}, we compare four baryonic properties: stellar mass, gas metallicity, gas mass, and black hole mass. 
First, in panel {\it (a)}, we observe that subhalos in TNG100-1 have $\lesssim 0.5$ dex higher stellar mass than those in TNG100-2 across the entire stellar mass range. 
The difference in stellar masses are attributed to the (subgrid) baryon physics models in TNG, which exhibits a higher star formation rate and stellar mass in a simulation with higher resolution \citep{2018MNRAS.473.4077P, 2018MNRAS.475..648P}. 
Our results align with the earlier studies in which the stellar-to-halo mass ratio was found to be a factor of two higher in TNG100-1 compared to TNG100-2 for halos with $M_{\rm DM} \sim 3 \times 10^{11} \msun$ \citep{2018MNRAS.473.4077P}.
Note that some subhalos which lack a stellar component in either of two simulations are excluded. 
In the upper marginal panel, the black solid line represents the fraction of subhalos of a given stellar mass in TNG100-1 that also have stellar particles in their TNG100-2 counterparts. 
One can see, for example, that less than half of the subhalos with $M_{\rm star} \sim 10^7 \msun$ in TNG100-1 have their counterpart subhalos in TNG100-2 that contain non-zero stellar particles. 
Conversely, the black dashed line in the right marginal plot indicates the fraction of subhalos in TNG100-2 that also have stellar particles in TNG100-1. Since it is easier for the subhalos at better spatial resolutions to host a gas cell that exceeds the star formation density threshold, many subhalo pairs have stellar particles only in TNG100-1.

The difference in star formation rates results in higher gas metallicities in the high-resolution versions of the two simulations, as depicted in panel {\it (b)} of Figure \ref{fig:mass_feat}. 
Significant scatters exist between the subhalo pairs, suggesting that the metallicity variations are influenced not only by the resolution differences, but also by factors beyond the resolution biases, such as the baryon cycle in the galaxies \citep{2008MNRAS.387..577O}. Here, the gas metallicity is calculated in a region within twice the stellar half-mass radius. 
It is noteworthy that for most of the subhalos in TNG100-1 with gas metallicities of $\log[Z_{\rm gas}/{\rm Z}_\odot] \lesssim -1.5$, their counterpart subhalos in TNG100-2 have predominantly null values in gas metallicity, due to resolution limit. 
In these halos, either the gas components are absent, or the stellar components that define the stellar half-mass radius are absent. 
We note that similar trends are found in panels {\it (a)} and {\it (b)} of Figure \ref{fig:mass_feat-50} when comparing TNG50-1 and TNG50-2. 

Next, in panel {\it (c)} of  Figure \ref{fig:mass_feat} we compare the gas masses of matching subhalos. 
While gas mass shows a relatively good agreement between the two simulations for subhalos with high gas mass ($M_{\rm gas} > 10^{10} \msun$), the gas mass in TNG100-2 tends to be higher than that in TNG100-1 for subhalos with $M_{\rm gas} \sim 10^{8.5} \msun$, seen as a ``kink'' in that mass range. 
For these halos the gravitational potential well is relatively shallow ($M_{\rm DM} \sim 10^{10} \msun$). 
Therefore, stellar feedback and other baryonic physics can expel gas from the gravitational potential well of the halo \citep{2015MNRAS.451.1247S}. 
The higher star formation rate and the associated increase in stellar feedback in the high-resolution simulation (TNG100-1) results in a lower gas mass than that in the low-resolution simulation (TNG100-2). 
To verify this scenario, with a blue dot-dashed line, we show the fraction of subhalos which have non-zero stellar masses in TNG100-1. And with an orange dotted line we plot the fraction of subhalos which have non-zero stellar masses in their TNG100-2 counterparts for a given $M_{\rm gas}$  in TNG100-1. 
Here one observes that in most subhalos of $M_{\rm gas} \sim 10^{8.5} \msun$, stars form only in TNG100-1, but not in TNG100-2. 
This entails a lack of stellar feedback in TNG100-2 that would have otherwise expelled the gas. Consequently, the gas mass of a subhalo in TNG100-2 tends to be higher than that in TNG100-1 at $M_{\rm gas} \sim 10^{8.5}\msun$. 
However, we note that this ``kink'' is less pronounced in panel {\it (c)} of Figure \ref{fig:mass_feat-50} where we compare TNG50-1 and TNG50-2.  
It is primarily due to the better numerical resolution of the lower-resolution simulation. 
The two simulations show relatively good convergence for $M_{\rm gas} \gtrsim 10^{7.5} \msun$. 

At the low-mass end of $M_{\rm gas} < 10^{8} \msun$, however, stars do not form in most subhalos in either TNG100-1 or TNG100-2. \my{Having an insufficient number of gas parcels ($\lesssim10$ parcels in TNG100-2), the gas mass of a subhalo in TNG100-2 tends to be lower than that in TNG100-1 at $M_{\rm gas} < 10^{8} \msun$, or the subhalo may even lack a gas component entirely.}\footnote{\my{While some studies with other numerical codes suggest that the baryon fraction of galactic halos decreases more significantly at the epoch of reionization as the mass resolution worsens \citep[e.g.,][]{2017MNRAS.467.1678Q}, reionization in the TNG model has marginal effects on the baryon masses of subhalos, regardless of resolutions \citep{2022MNRAS.512.4909G, 2023MNRAS.525.5932B}.}}

Lastly, there is a resolution-wide agreement in the mass of MBH of a subhalo as can be seen in panel {\it (d)} of Figures \ref{fig:mass_feat} and  \ref{fig:mass_feat-50}. It shows similar black hole masses in both the high- and low-resolution subhalos, except when near the resolution limit of the low-resolution simulation, $M_{\rm BH} \sim 10^{7.5} \msun$ (for TNG100-2) and $\sim 10^{6.5} \msun$ (for TNG50-2). 
For subhalos harboring an MBH, most of their counterparts in the low-resolution simulation also have an MBH, unlike the trend observed in gas and stellar components.  
This can be seen in the marginal plot at the top of panel {\it (d)}, which shows the fraction of MBH-harboring subhalos in TNG100-1 (TNG50-1) for which an MBH is also present in its matching halo in TNG100-2 (TNG50-2).  
This behavior can be attributed to the TNG physics model that implements an MBH seed for all halo above a certain mass threshold \citep{2017MNRAS.465.3291W}. 
Since the seeding mechanism depends solely on the DM properties --- which show a great convergence as illustrated in Figures \ref{fig:halo_feat} and \ref{fig:halo_feat-50} --- the black hole mass in the TNG simulations exhibit robust numerical convergence with resolution. 
We caution the readers about the difference between TNG100 (Figures \ref{fig:mass_feat}) and TNG50 (Figures \ref{fig:mass_feat-50}) observed at the low-mass end. 
While subhalos in TNG100-1 tend to harbor less massive MBHs than those in TNG100-2 do, subhalos in TNG50-1 tend to have more massive MBHs than those in TNG50-2 do. 
This is because TNG100-1 and TNG100-2 use different MBH seed masses ($M_{\rm seed}$), whereas TNG50-1 and TNG50-2 adopt the same seed mass, $M_{\rm seed} = 8\times 10^5 \msunh$ \citep{2017MNRAS.465.3291W}.

\begin{figure*}
    \centering
    \includegraphics[width=0.9\linewidth]{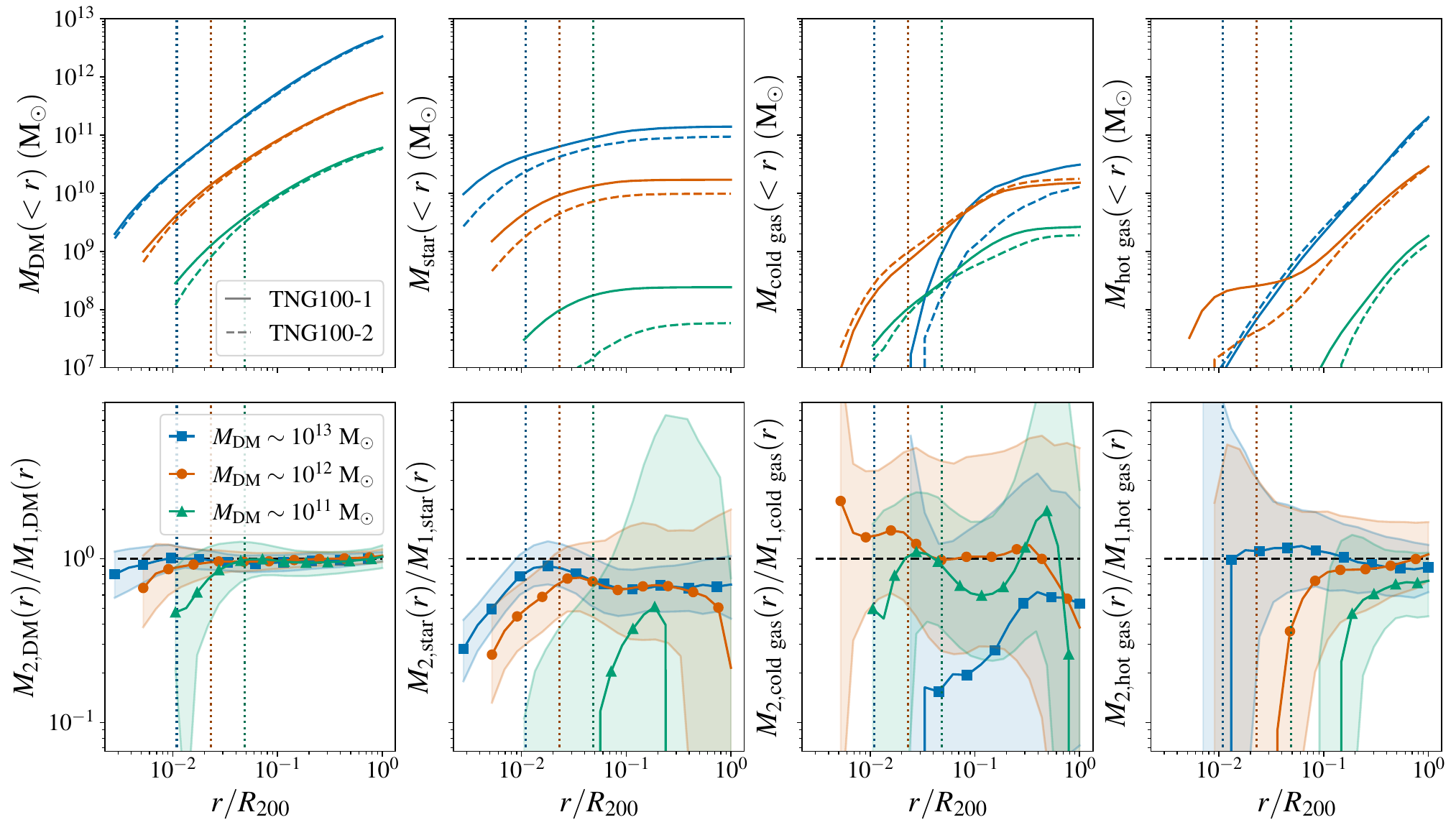}
    \caption{{\it Top row}: Median enclosed-mass profile of ``central’’ galaxies in various sample mass ranges at $z=0$.    
    Solid lines denote subhalos in TNG100-1 (high-resolution run), while dashed lines represent those in TNG100-2 (low-resolution run). 
    {\it From left to right}, the median enclosed mass profiles for the DM, stellar, cold gas, and hot gas components are depicted, respectively. 
    The {\it bottom row} presents the median values of the mass ratios in each radius bin (dividing TNG100-2 by its TNG100-1 counterpart). The shaded regions indicate $16^{\rm th}-84^{\rm th}$ percentile ranges. 
    {\it Vertical dotted lines} in each panel show the three times the gravitational softening length of TNG100-2 with respect to mean virial radius of each mass bin, denoting the resolution limit of the lower-resolution simulation in each mass bin. 
    In general, massive subhalos show better agreement with resolution changes, but there is a noticeable deficit of cold gas in the low-resolution galaxies with $M_{\rm DM}\sim 10^{13}\msun$. 
    For a detailed discussion on this plot, see Section \ref{sec:result_rc}. }  
    \label{fig:rc_ratio}
    \vspace{1mm}
\end{figure*}

\subsection{Galaxy-by-galaxy Resolution Study On Radial Profiles of Halos}\label{sec:result_rc}
After compiling the subhalo-matching catalog between two simulations of different resolutions, we are able to conduct a galaxy-by-galaxy comparison study on the subhalos' radial profiles.
 In Figure \ref{fig:rc_ratio}, we examine the differences in the spatial distribution of particles within subhalos by plotting the radial profiles of enclosed masses in TNG100-1 and TNG100-2 (top row).\footnote{We adopt the positions of the particles with minimum gravitational potential energy as the centers of the galaxies.} 
We also plot the mass ratios in each radius bin (bottom row; dividing TNG100-2 by TNG100-1). 
We group the subhalos into three mass bins in units of $M_{\rm DM}/\msun$ as [$10^{12.5}, 10^{13.5}$], [$10^{11.5}, 10^{12.5}$], and [$10^{10.5}, 10^{11.5}$]. Each is denoted in the figure as $\sim 10^{13}$, $\sim 10^{12} $, and $\sim 10^{11} \msun$, respectively. We also limit our analysis to central galaxies, which is the primary subhalo in its FoF group. Subhalos that possess no stellar particles, constituting 10\% of the lowest mass bin, are excluded from this analysis. 
For reference, in each panel, we display vertical dotted lines representing three times the gravitational softening lengths of TNG100-2, divided by the mean virial radius of the halos in each mass bin, $3 \times \epsilon_{\rm DM,\,star}^{z=0}/\langle R_{200}\rangle$ \citep{2019MNRAS.490.3196P}.\footnote{The softening length of collisionless particles in TNG100-1 is a half of that in TNG100-2, being 0.74 kpc and 1.48 kpc, respectively (see Table \ref{tab:tng}).}

In the first column of the figure, the DM mass shows strong convergence between the two simulations, except near the resolution limit (\my{$r/R_{200} \lesssim 0.01,\ 0.02,\ 0.05$ for each mass bin, or $r\lesssim 4$ kpc}), where the density in the low-resolution simulation is approximately 60\% of that in the high-resolution simulation. This result is in line with previous studies, such as \cite{2000ApJ...529L..69J}. 
The baryon components such as stars and gas display more significant differences than what is seen in the DM profiles. 
In the second column, the deficit in stellar mass in the low-resolution simulation is more pronounced in the inner region, \my{within a few times of the gravitational softening length, compared to that in DM mass}. 
Moreover, the median ratio in the outer region situated $\sim 0.6$, showing clear deficit of stellar particles in the low-resolution galaxies with $M_{\rm DM} \sim 10^{13}$ and $\sim 10^{12}$. This trend can be observed more clearly in the lowest mass bin, $M_{\rm DM} \in [10^{10.5}, 10^{11.5}]\msun$, where the median value of the stellar profile ratios is lower than 0.4 at all radii. 

We label gas parcels with $T<10^{4.5}\,{\rm K}$ as {\it cold gas} and the remainder as {\it hot gas}, following \cite{2023MNRAS.524.3502R}. For galaxies with $M_{\rm DM} \in [10^{12.5}, 10^{13.5}]\msun$, the median ratio of the cold gas masses between the two simulations is below unity at all radii, despite a large variance (see the third column of Figure \ref{fig:rc_ratio}). The deficit of gas is more pronounced near the central galaxies ($r/R_{200} \lesssim 0.2$) whereas the circumgalactic medium ($r/R_{200} \gtrsim 0.2$) shows better convergence. In contrast, galaxies with $M_{\rm DM} \sim 10^{12}\msun$ and $\sim 10^{11}\msun$ exhibit better agreement with the resolution change; their median cold gas mass ratio is close to unity. We observe the opposite result for the hot gas distribution (fourth column). The galaxies in the highest mass bin ($M_{\rm DM} \sim 10^{13}$) show good agreement with the resolution change, but those in lower mass bins tend to have decreased hot gas density near the central galaxies ($r/R_{200} \lesssim 0.1$ for the galaxies with $M_{\rm DM} \sim 10^{12}\msun$ and $r/R_{200} \lesssim 0.3$ for galaxies with $M_{\rm DM} \sim 10^{11}\msun$) in the low-resolution simulation.\footnote{The steep decline in the median hot gas ratio at $r/R_{200} \sim 0.01$ for galaxies with $M_{\rm DM} \sim 10^{13}\msun$ is induced by an insufficient number of hot gas parcels.} In the highest mass bin, AGN feedback from supermassive black holes plays important role in suppressing the cold gas mass \citep{2020MNRAS.497..146D}. We speculate that gas content in the low-resolution is more easily heated by AGN feedback, as reported by \cite{2015MNRAS.453.1829B}. In the lower mass bins, where stellar feedback dominates over AGN feedback, the reduced stellar masses in the low-resolution galaxies could lead to the decrease in hot gas content in the inner region.

\begin{figure*}
    \vspace{0mm}
    \centering
    \includegraphics[width = 1.005\linewidth]{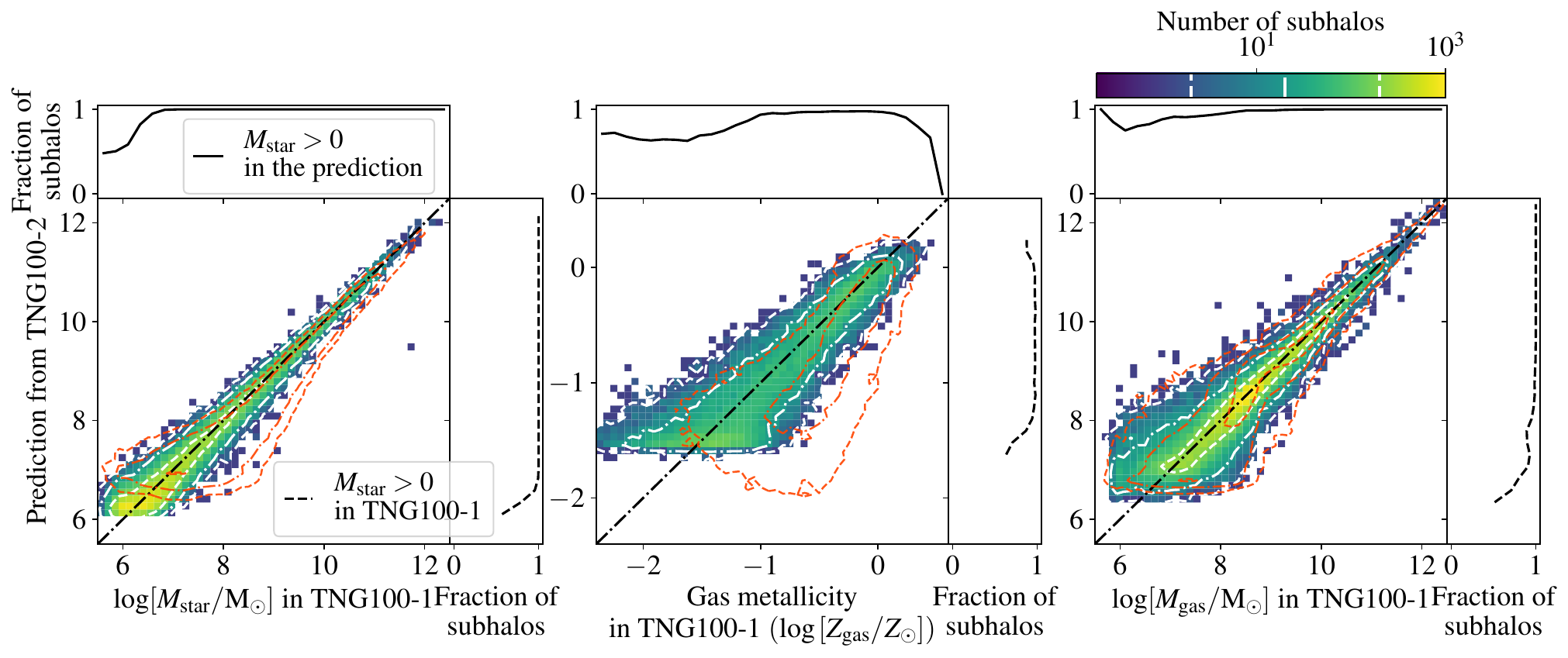}    
    \vspace{-3mm}    
    \caption{The comparison of baryonic subhalo properties between the true values in TNG100-1 (high-resolution run) and the machine-predicted values based on the properties in TNG100-2 (low-resolution run).  
    Our ML model predicts a subhalo's stellar mass, gas metallicity, and gas mass in the high-resolution simulation TNG100-1 ({\it from left to right}), based solely on its matching pair's properties in the low-resolution simulation TNG100-2.  
    Using the ML technique, we successfully ``correct'' the resolution biases exhibited in Figure \ref{fig:mass_feat} (notice that the white contours in Figure \ref{fig:mass_feat} are copied to this figure as {\it red contours}). 
    The range of the color bar is scaled down from that of Figure \ref{fig:mass_feat}, by the ratio of the number of subhalos in the test set to the total number of subhalos. 
    See Section \ref{sec:ml_result} for more details.}
    \label{fig:ml_gas}
    \vspace{2mm}
\end{figure*}

   \begin{table*}
   \centering
   \begin{tabular}{lc|cccc}
       \toprule
       Target & Method & RMSE & MAPE & Bias & Pearson $\rho$ \\
       \midrule
       \multirow{3}{*}{\textit{Stellar mass}} & Before correction (Figure \ref{fig:mass_feat}) & 0.659 & 0.069 & 0.4959& 0.921\\
        & After correction by gaussian process regression & 0.417 & 0.040 & $0.0000$ & 0.928 \\
        & After correction by ML (Figure  \ref{fig:ml_gas}) & 0.204 & 0.019 & 0.0063 & 0.983\\
       \midrule
       \multirow{3}{*}{\textit{Gas metallicity}} & Before correction (Figure \ref{fig:mass_feat}) & 0.506 & 0.190 & 0.3720 & 0.729\\
        & After correction by gaussian process regression & 0.260 & 0.086 & 0.0002 & 0.751\\
        & After correction by ML (Figure  \ref{fig:ml_gas}) & 0.175 & 0.059 & 0.0073 & 0.897\\
       \midrule
       \multirow{3}{*}{\textit{Gas mass}} & Before correction (Figure \ref{fig:mass_feat}) & 0.477 & 0.043 & 0.1877 & 0.902\\
        & After correction by gaussian process regression & 0.364 & 0.031 & -0.0000 & 0.920\\
        & After correction by ML (Figure  \ref{fig:ml_gas}) & 0.269 & 0.022 & -0.0025 & 0.957\\
       \bottomrule
   \end{tabular}
   \vspace{1mm}   
   \caption{Comparison of performance metrics for three subhalo properties: stellar mass, gas metallicity, and gas mass.  
   Here we compare the root mean squared errors (RMSE), the mean absolute percentage errors (MAPE), bias, and Pearson correlation coefficient ($\rho$) when comparing the matching subhalos' properties between TNG100-1 and TNG100-2. The first method ``before correction'' means the pure, inherent resolution effect seen in Figure \ref{fig:mass_feat}, between two simulations of different resolutions.  
   The second method is the error after we have applied a {\it gaussian process regression} model to compensate for the resolution biases.  
   Errors in stellar mass and gas metallicity are significantly reduced when the gaussian process regression model is applied. 
   The ML model further reduces the error across all three properties, as illustrated in Figure \ref{fig:ml_gas}.
   See Section \ref{sec:ml_result} for more details.}
   \label{tab:model_comparison}
   \vspace{1mm}
\end{table*}

\subsection{Mitigating the Resolution Biases: Correcting Low-resolution Subhalos With Machine Learning}\label{sec:ml_result}

ML techniques have been used to paint baryonic properties onto DM halos in DMO simulations after it learned the relationship between the DM properties and baryonic properties of galactic halos \citep[e.g.,][]{2019MNRAS.489.3565J,2022MNRAS.509.5046L,2022ApJ...941....7J}. 
Inspired by these pioneering studies, we aim to ``correct'' the baryonic properties of halos found in low-resolution simulations, by constructing a regression model that learns the relationship between the halo properties in a low-resolution simulation and in a high-resolution simulation (see Section \ref{sec:correcting}).
In this section, we illustrate the performance of the ML model we have built.  
Then, as a first test, we apply the ML model to the subhalos in TNG300-1, which ``corrects'' their baryonic properties to match those in a higher resolution benchmark, TNG100-1. 

\begin{figure*}
    \centering
    \includegraphics[width = 0.75\linewidth]{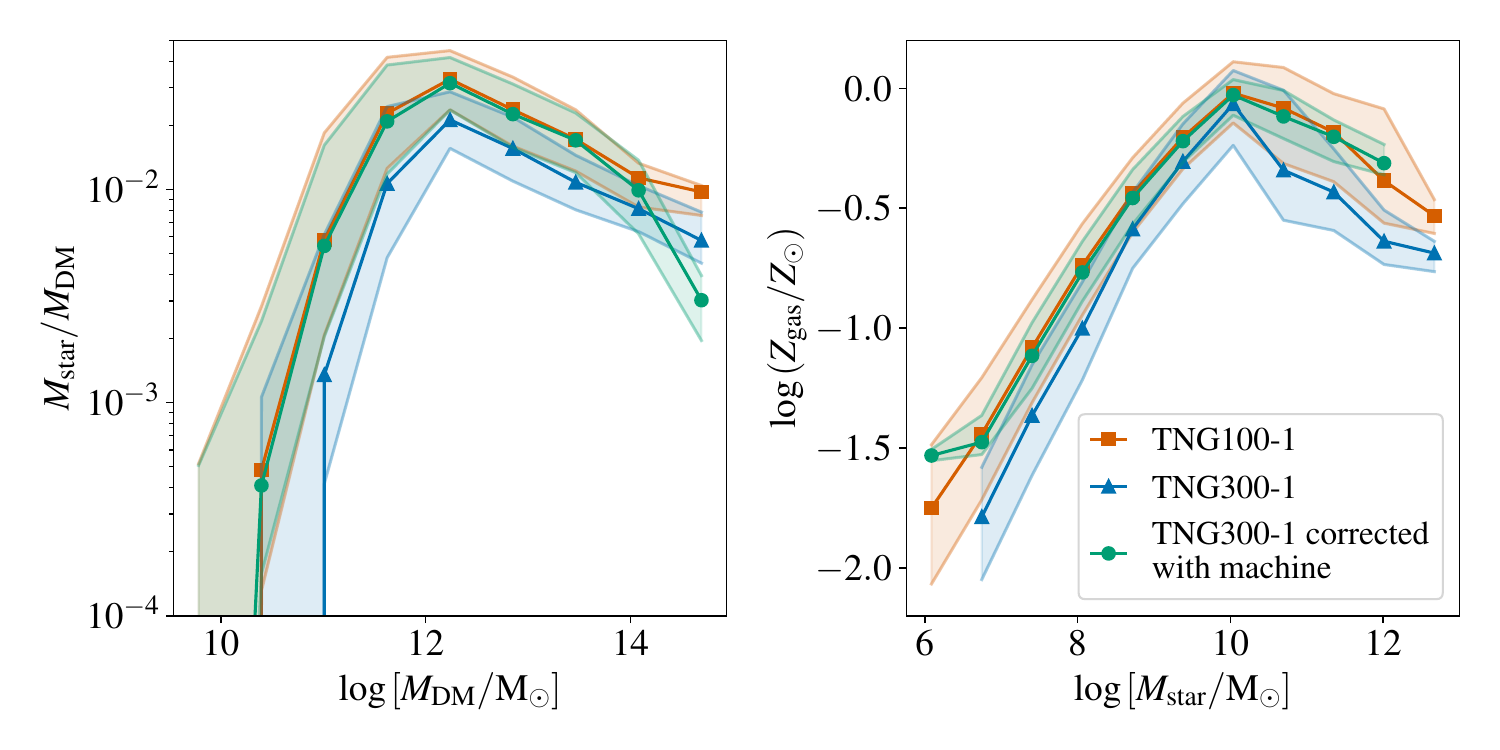}
    \vspace{-3mm}    
    \caption{Correction of subhalo properties at $z=0$ in TNG300-1 (low-resolution run) using an ML model trained on TNG100-1 (high-resolution run) -- TNG100-2 (low-resolution run) subhalo pairs. 
    The {\it left panel} displays the stellar mass -- halo mass (SMHM) relation for subhalos in TNG100-1 and TNG300-1, and for the TNG300-1 subhalos after ML-based corrections.  
    The {\it right panel} shows the stellar mass -- gas metallicity relation (MZR) for the same subhalos. 
    Shaded areas indicate the $16^{\rm th}-84^{\rm th}$ percentile ranges. 
    While a systematic discrepancy exists between TNG300-1 ({\it blue triangles}) and TNG100-1 ({\it orange squares}), the corrected TNG300-1 subhalos ({\it green circles}) align very closely with the TNG100-1 subhalos, both in the SMHM relation and in the MZR. 
    Interestingly, the ML model reproduces the scatter of the SMHM relation seen in the TNG100-1 subhalos (i.e., the green-shaded region nearly overlaps the orange-shaded region). 
    See Section \ref{sec:ml_result} for more details.}
    \label{fig:ml_tng300}
    \vspace{2mm}
\end{figure*}

In the left panel of Figure \ref{fig:ml_gas}, we present the machine's predictions of subhalo stellar masses.
The machine predicts a subhalo stellar mass in a high-resolution simulation TNG100-1, solely based on its matching pair's stellar mass in a low-resolution simulation TNG100-2.  
Our ML model is shown to predict a subhalo's stellar mass in TNG100-1 with great accuracy, using only the information from TNG100-2. 
This is remarkable especially when compared to Figure \ref{fig:mass_feat} (or red contours in Figure \ref{fig:ml_gas}), which clearly demonstrates the systematic discrepancy between high- and low-resolution simulations.  
Notice that the white contours in Figure \ref{fig:mass_feat} are copied to Figure \ref{fig:ml_gas} as red contours in order to guide the eyes and compare the two figures. 
Similar to Figure \ref{fig:mass_feat}, in the upper marginal plot, the black solid line represents the fraction of subhalos of a given stellar mass in TNG100-1 that also have stellar particles in their ``corrected'' TNG100-2 counterpart halos. 
For example, nearly all the subhalos with $M_{\rm star} \sim 10^7 \msun$ in TNG100-1 have their ``corrected'' counterpart subhalos in TNG100-2 containing some stellar particles.  
This finding is in stark contrast to Figure \ref{fig:mass_feat}, where only less than half of these TNG100-1 halos had counterpart TNG100-2 subhalos with any stars.  
While a low-resolution simulation like TNG100-2 tends to lack stellar particles towards the lower end of the stellar mass range, our prediction model is shown to be able to ``correct'' the stellar masses of these subhalos across nearly the entire range of stellar masses above $\sim 10^7 \msun$.

To quantitatively evaluate the performance of our model, in Table \ref{tab:model_comparison} we present the root mean squared error (RMSE), mean absolute percentage error (MAPE), bias, and Pearson correlation coefficient ($\rho$), when comparing the matching subhalos' properties between TNG100-1 and TNG100-2.\footnote{Errors are calculated for all the subhalos for which both the stellar mass predicted from TNG100-2 and the actual stellar mass in the two simulations are nonzero. For these calculations we use logarithmic values, $y = \log{[M_{\rm star}/\msun]}$. The same approach is adopted for the gas metallicity (middle panel of Figure \ref{fig:ml_gas}) and gas mass (right panel).  Although we represent gas metallicity using the unit $\log{[Z_{\rm gas}/Z_{\odot}]}$,  we compute the metrics using $\log{[Z_{\rm gas}]}$ to prevent MAPE from diverging when $\log{[Z_{\rm gas}/Z_{\odot}]}$ is close to zero.} 
The first column ``before correction'' means the pure, inherent error due to the resolution biases between TNG100-1 and TNG100-2, as can be clearly seen in Figure \ref{fig:mass_feat}. 
The second column is the error after we have applied a {\it gaussian process regression} model to compensate for these resolution biases.\footnote{This simply shifts the stellar mass values by using only the stellar mass from the low-resolution simulation as its input. After the correction, biases become close to zero (see Table \ref{tab:model_comparison}).} One can see that the errors in stellar mass are greatly reduced even with this simple fix. 
Then, the third column shows that our ML model further reduces the \my{RMSE and MAPE} in stellar mass by $\sim 50\%$, as is seen in Figure \ref{fig:ml_gas}.

For the gas metallicity in the middle panel of Figure \ref{fig:ml_gas}, we again see that our ML model successfully ``corrects'' the properties of low-resolution subhalos to match those of the high-resolution counterparts.  
However, the scatter is larger compared to that  in the stellar mass panel, with a MAPE value about three times as high as that for stellar mass (0.059 vs. 0.019). The predictions also exhibit a steep lower bound at $\log{[Z_{\rm gas}/Z_{\odot}]} \sim -1.5$, indicating that the model complexity is significantly lower compared to those of other models.   
In addition, about 30\% of the subhalos with $\log{[Z_{\rm gas}/Z_{\odot}]} < -1.5$ in TNG100-1 have their ``corrected'' counterpart subhalos in TNG100-2 with a wrongly-predicted, null metallicity value (see the upper marginal plot). 
The metallicity prediction is particularly challenging because there is a large scatter in the training set (see panel  {\it (b)} in Figure \ref{fig:mass_feat}), and because most TNG100-1 subhalos with $\log{[Z_{\rm gas}/Z_{\odot}]} < -1.5$ have their counterpart TNG100-2 subhalos with null gas metallicity due to its resolution limit (see the upper marginal plot of panel  {\it (b)} in Figure \ref{fig:mass_feat}). 
Yet, the machine still performs significantly better than a gaussian process regression, reducing \my{the RMSE and MAPE by $\sim 33\%$} (see Table  \ref{tab:model_comparison}).

Our ML model for gas mass also performs well, as shown in the right panel of Figure \ref{fig:ml_gas}.  
Most notably, the ``kink'' seen in panel {\it (c)} of Figure \ref{fig:mass_feat} has largely disappeared.  
The performance slightly deteriorates compared to the stellar mass model, exhibiting a larger RMSE and MAPE in Table  \ref{tab:model_comparison}. 
In the right marginal plot, the fraction, which denotes the accuracy of the model, deviates from unity for subhalos with $M_{\rm gas} \lesssim 10^8 \msun$. 
Nevertheless, when compared to the \my{gaussian process regression} approach, the ML model reduces the \my{MAPE by $\sim 30\%$} (see Table  \ref{tab:model_comparison}). 

We conclude that ML techniques can help predict the properties of a subhalo in a high-resolution simulation based solely on its properties in a low-resolution simulation.
The prediction is particularly good for the stellar mass and gas mass of subhalos, thereby mitigating resolution biases. \my{In Appendix \ref{sec:ml_sate}, we present results for both central and satellite subhalos, which show no notable difference in performance, except in the prediction of gas mass. }

Now we apply this ML model to the subhalos in a large cosmological simulation volume (TNG300-1) and ``correct'' their properties to match those in a higher resolution benchmark, TNG100-1. 
TNG300-1 has an identical resolution as TNG100-2, but with a larger ($302.6$ cMpc$)^3$ box size. 
Therefore, we can apply our trained ML model to the subhalos in TNG300-1, adjusting their subhalo properties to those in TNG100-1. 
In Figure \ref{fig:ml_tng300}, we show two important relationships --- the stellar mass -- halo mass (SMHM) relation and the stellar mass -- gas metallicity relation (MZR) --- from the two flagship TNG simulations with different box sizes.\footnote{The total DM mass found in the {\sc SubFind} algorithm (see Section \ref{sec:tng}), {\tt SubhaloMassType1}, is used as a proxy for the subhalo (galaxy) mass.}
First, it can be observed that a systematic discrepancy exists between TNG300-1 (blue triangles) and TNG100-1 (orange squares) due to the fact that the resolution of TNG300-1 is 8 times worse in mass compared to that of TNG100-1.   
The deficit in stellar mass in a low-resolution simulation is exactly what is shown in panel {\it (a)} of Figures \ref{fig:mass_feat} and \ref{fig:mass_feat-50}.  
Since both the stellar mass and the gas metallicity of a subhalo tend to increase with resolution (i.e., panels {\it (a)} and {\it (b)}  of Figures \ref{fig:mass_feat} and \ref{fig:mass_feat-50}), a higher resolution simulation TNG100-1 yields a slightly higher MZR than TNG300-1 does. 

Second, the green circles in Figure \ref{fig:ml_tng300} display the results from our ML model where properties of the subhalos in TNG300-1 are ``corrected'' as if they are from a higher resolution simulation. 
The SMHM relation of these ``corrected'' subhalos (green circles) in the left panel becomes nearly identical to that in TNG100-1 (orange squares), except on the high-mass end ($M_{\rm DM} > 10^{14}\,M_{\odot}$), where the amount of the training set is limited due to a smaller (106.5 \text{cMpc})$^3$ box size. 
The ``corrected'' MZR also closely matches that of TNG100-1, except on the high- and low-mass ends.
Interestingly, the ML model is shown to be able to reproduce the same scatter in the SMHM relation, with a similar 1-sigma shade (68\% confidence interval) as the TNG100-1's relation (i.e., the green-shaded region overlaps the orange-shaded region). 
Achieving this has been often considered unlikely when only the features from a DMO simulation were used as inputs to an ML model \citep{2018MNRAS.478.3410A}. 
However, because we directly provide information on baryonic properties, the model does not heavily rely on their halo mass at $z=0$ alone, but also uses various other inputs to replicate the galaxy-to-galaxy scatter seen in a hydrodynamic simulation. 
And yet, we fail to reproduce the same amount of scatter when predicting the MZR. 
This finding is in agreement with the ML model's relatively weak performance in predicting metallicity, as demonstrated in Figure \ref{fig:ml_gas}.

\section{Discussion}\label{sec:discussion}

\subsection{Further Verification of the Matching Algorithm}\label{sec:verifi}

We have demonstrated that the subhalo in our matching catalog have nearly identical peculiar velocities, as well as similar DM masses and velocity dispersions in the high-resolution simulation, TNG100-1 and TNG50-1, and its low-resolution counterpart, TNG100-2 and TNG50-2 (see Sections \ref{sec:matching} and \ref{sec:result_halo}). 
In this section, we further test our matching algorithm by tracking the DM particles of the matching subhalo pairs. Although we only identify the matching pairs based only on the positions of subhalos along their evolution histories (see Sections \ref{sec:matching}), it is expected that DM particles in these subhalo pairs indeed come from the same Lagrangian patch in the initial conditions. 
We trace each DM particle of the subhalo in TNG100-1 back to the initial condition at $z=127$, and locate the corresponding particles by finding the nearest particles in position in the initial condition of TNG100-2. 
Finally, we check whether these corresponding particles indeed belong to the matched TNG100-2 subhalo in our matching catalog. 

\begin{figure}
    \centering
    \vspace{2mm}
    \includegraphics[width = 0.93\linewidth]{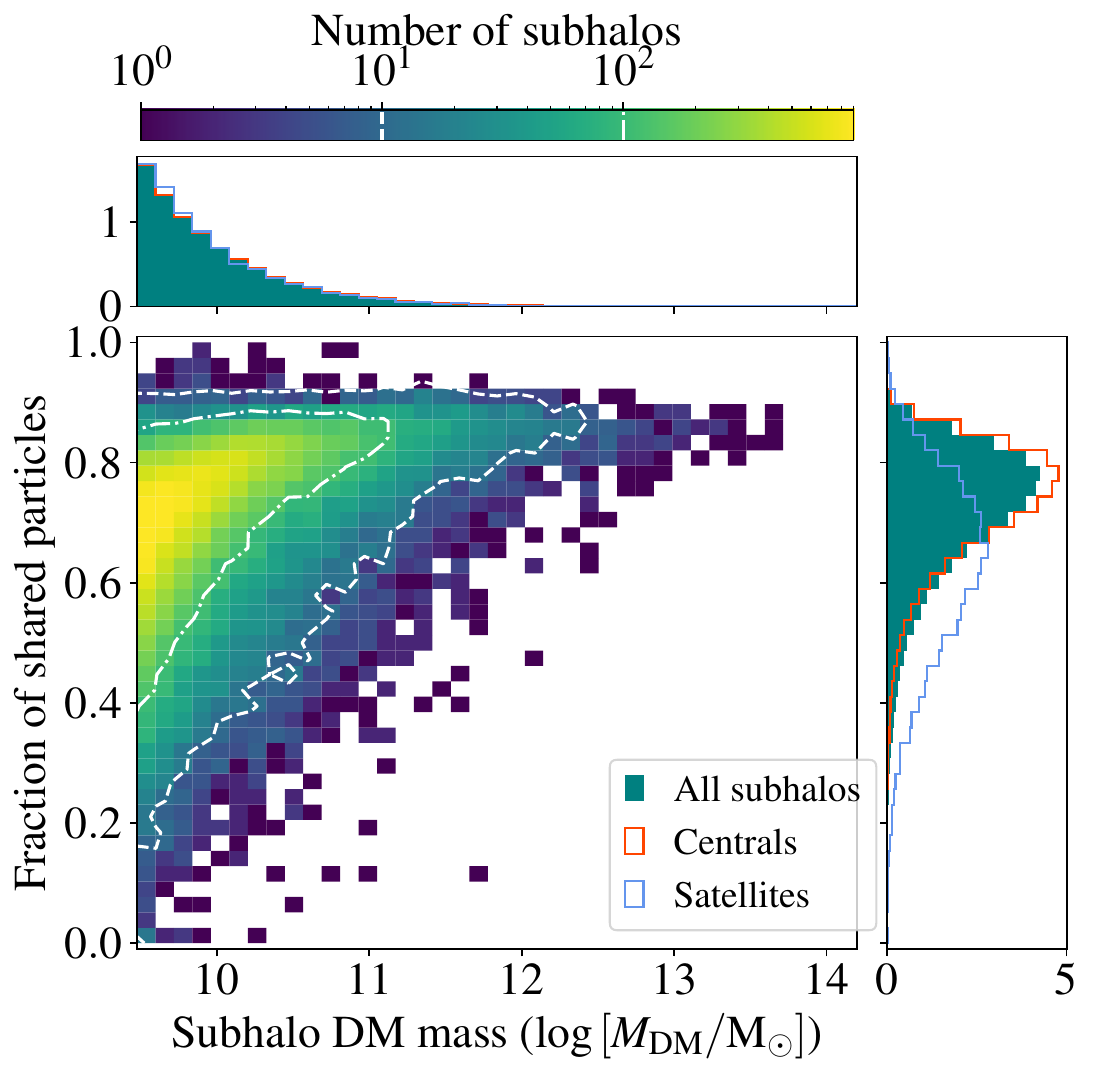}
    \caption{The fraction of DM particles in a TNG100-1 subhalo that have corresponding particles in its matching subhalo in TNG100-2. 
    A fraction of 1.0 suggests that all DM particles in a TNG100-1 subhalo at $z=0$ are present in its counterpart TNG100-2 subhalo. 
    We plot only the subhalos that exceed the mass threshold for the matching process, $3 \times 10^9 \msun$ (see Section \ref{sec:matching}). 
    Due to computational constraints, only \my{50,000} subhalos are represented in this plot. 
    \my{The {\it marginal plots} on the right and at the top display density histograms, showing that most matched pairs share approximately 70\% of DM particles. Orange and blue histograms represent central and satellite subhalos, respectively, with the satellites exhibiting lower fractions.} See Section \ref{sec:verifi} for more details.}    
    \vspace{1mm}
    \label{fig:predict_verif}
\end{figure}

In the left panels of Figure \ref{fig:predict_verif}, we present the fraction of DM particles in a TNG100-1 subhalo that have corresponding particles inside its matching subhalo in TNG100-2. 
For the practical purpose, we limit our investigation to a subset of \my{50,000} subhalos. 
The fractions are plotted against the smaller of the two DM masses of each matched subhalo pair. 
For each matched pair of subhalos, we take the DM mass that is smaller and use that value for the plot.
On average, a TNG100-1 subhalo shares 70\% of its DM particles with its counterpart halo in TNG100-2. \my{Based on the fractions of the pairs from the catalog provided by the TNG Collaboration, we consider pairs with a fraction below 0.4 to be insufficient for matching (see Figure \ref{fig:predict_verif_dmo}).}

\my{The fractions of satellite subhalos are lower than those of central subhalos (orange and blue density histograms in the marginal plots of Figure \ref{fig:predict_verif}). This finding aligns with the limited accuracy of our matching method for satellites during the verification step (see Figure \ref{fig:matching_hist_lowres}). Furthermore, 0.89\% of centrals and 8.86\% of satellites (2.38\% in total) have a fraction below 0.4, which is considered insufficient for a match in the matching method based on particle IDs. Therefore, it is apparent limitation of our study that including those subhalos in the analysis. Nevertheless, only 0.024\% of the subset have a fraction of 0.0, indicating that the two matched pairs originated from completely unrelated patches in the initial conditions. } 
 
\my{The subhalo pairs with an insufficient number of shared DM particles complicate the `galaxy-by-galaxy' comparison approach. The dark matter halos of these pairs are too dissimilar to effectively study the impact of resolution changes on baryon physics. Furthermore, the chaotic behavior inherent in cosmological simulations introduces intrinsic scatter between the simulations; galaxies and circumgalactic media can diverge from each other by a tiny shift in the early universe, even for the pairs with sufficient shared DM particles \citep{2019ApJ...871...21G, 2019MNRAS.482.2244K, 2023MNRAS.526.2441B}. These studies show that minute perturbations in the early universe can lead to galaxy-level differences in cosmological simulations, especially in the presence of feedback. Late major mergers amplify this stochasticity \citep{2019MNRAS.482.2244K}, making satellites in dense environments particularly challenging to compare. Similarly, \cite{2021MNRAS.507.4953G} show that the trajectories of matched satellites at two different resolutions can diverge after passing the second pericentric passage. Consequently, satellite subhalos in different simulations could be highly dissimilar. Therefore, comparing subhalos across simulations necessitates examining a substantial number of pairs to draw meaningful conclusions. A conclusion drawn from a single galaxy pair is susceptible to being severely impacted by these chaotic effects.}  

We have demonstrated that \my{most of} the subhalo pairs originating from almost the identical regions in the initial conditions can be identified by only using the mass and positions of main progenitors. \my{In Appendix \ref{sec:dmo_comparison}, we compare the fractions with those obtained from the subhalo pairs identified between DMO and hydrodynamic simulations. These fractions are higher than those in comparisons between TNG100-1 and TNG100-2, which suggests that the resolution difference has a greater impact on the similarity between the subhalo pairs than the presence of baryons.}

\begin{figure*}
    \vspace{1mm}
    \centering
    \includegraphics[width = \linewidth]{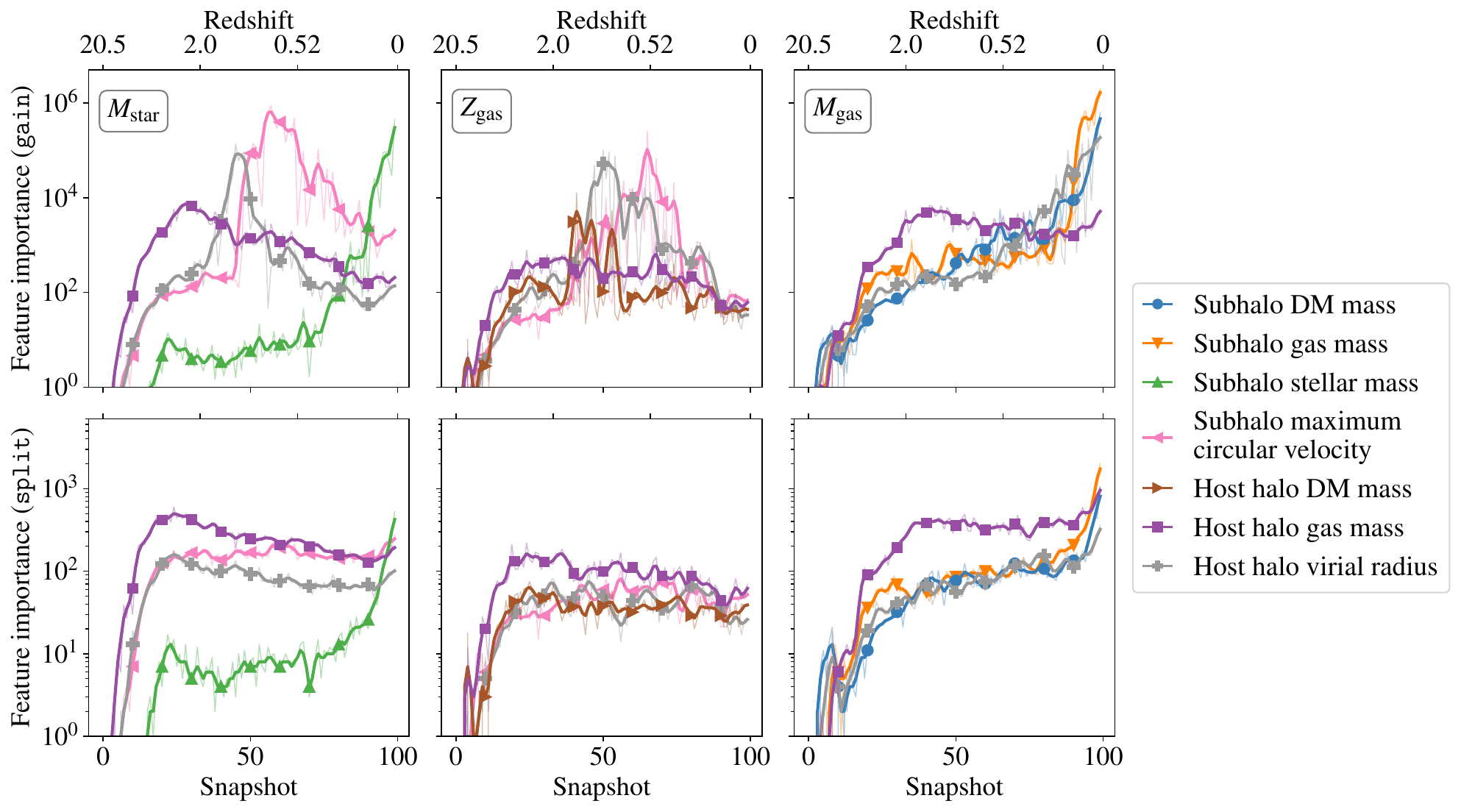}
    \vspace{-1mm}    
    \caption{Feature importances of our ML model that predicts a subhalo's stellar mass, gas metallicity, and gas mass in the high-resolution simulation TNG100-1 ({\it from left to right}), based on its matching pair's properties in the low-resolution simulation TNG100-2 (Sections \ref{sec:correcting}, \ref{sec:ml_result}, and Figure \ref{fig:ml_gas}).  
    The number shows the degree to which the model depends on each input feature.
    The thin, semi-transparent lines show the raw data, while the thick lines represent the smoothed data. 
    We present two types of feature importances: \texttt{gain} ({\it top row}) and \texttt{split} ({\it bottom row}). 
    Since we provide a total of 2300 features (23 features $\times$ 100 snapshots) as inputs for a single subhalo, the feature importances can be computed for each of 23 features at each of 100 epochs.  
    The computed numbers are presented for the four features with the highest \texttt{gain} values in each prediction model at 100 epochs. 
    Notably, when predicting the stellar mass and gas metallicity of subhalos, the model heavily relies on the information from $z\gtrsim1$. 
    See Section \ref{sec:feature_importance} for more details.}
    \label{fig:feature_importance}
    \vspace{2mm}
\end{figure*}

\subsection{Feature Importances and the Ablation Study}\label{sec:feature_importance}

We have shown that an ML technique {\sc LightGBM}, a modified version of the gradient tree boosting algorithm, can be used to mitigate the resolution biases by correcting the subhalo properties in a low-resolution simulation to match those in a high-resolution simulation (see Sections \ref{sec:correcting} and \ref{sec:ml_result}). 
An inherent advantage of using decision tree-based regression models such as {\sc LightGBM} is the ability to identify the key attributes that significantly influence the prediction of the output, among numerous input features. 
In contrast to a deep learning model, where it is often a challenge to understand the specific procedures that led to a particular decision, the ability to explain the results is an advantage of a {\sc LightGBM} model.
This allows us to study which features are important to determine target properties. 
We study two types of feature importance of our ML model in Figure \ref{fig:feature_importance}, \texttt{gain} (top row) and \texttt{split} (bottom row). 
\texttt{Gain} represents the improvement in accuracy (or the reduction in loss) brought about by a feature. On the other hand, \texttt{split} indicates how many times the feature is used for decision-making within the gradient boosting decision tree. 
Since we feed a total of 2300 features (23 features $\times$ 100 snapshots) as inputs to the ML model, the feature importances can be computed for each of 23 features at each of 100 epochs.  
Hence, Figure \ref{fig:feature_importance} shows the computed importances of the four features with the highest \texttt{gain} values in each prediction model at 100 epochs. 

In the left column of Figure \ref{fig:feature_importance}, one can see that in order to predict the stellar mass of a subhalo in TNG100-1, the ML model relies --- obviously --- on the stellar mass of its counterpart subhalo in TNG100-2 at $z=0$ (\texttt{SubhaloMassType4} in Table \ref{tab:features}; green triangles).  
From $z\sim0.5$ to $z=0$, the machine gives increasing weight to the stellar mass, considering it as a crucial component for predictions. 
However, other features also contribute to the stellar mass prediction.  
For example, the \texttt{gain} values of the subhalo's maximum circular velocity  (\texttt{SubhaloVmax} in Table \ref{tab:features}; pink triangles) or the host halo's virial radius (\texttt{Group\_R\_TopHat200}; gray crosses) are close to or above $\sim10^5$ at $z\sim1$.  
Even more interestingly, the \texttt{split} value of the host halo's gas mass at $z\sim3$  (\texttt{GroupMassType0}; purple squares) is equal to or greater than  that of the subhalo's stellar mass at $z=0$.  
Our ML model seems to reflect the baryonic physics at cosmic noon, when the cosmic star formation rate reached its peak due to mergers and other galaxy-galaxy interactions. 

Given that gas metallicities are closely tied to stellar masses, the feature importances for the gas metallicity prediction model show similar results, as seen in the middle column of Figure \ref{fig:feature_importance}. 
The model again emphasizes the subhalo's dynamical mass  (\texttt{SubhaloVmax}; pink triangles) and the host halo's virial radius (\texttt{Group\_R\_TopHat200}; gray crosses) at high $z$. 
Since there exists a large scatter when comparing the gas metallicities between the low- and high-resolution simulations (see panel {\it (b)} of Figures \ref{fig:mass_feat} and \ref{fig:mass_feat-50}), the model does not rely as much on the the subhalo's metallicity in TNG100-2 to predict that in TNG100-1.  

In contrast to the two previous models, for the gas mass prediction, the ML model prefers to use features closer to $z=0$.
It bases its predictions on the subhalo's gas mass (\texttt{SubhaloMassType0}; orange triangles) and DM mass (\texttt{SubhaloMassType1}; blue circles), and its host halo's gas mass (\texttt{GroupMassType0}; purple squares).  
This suggests that the gas mass of the subhalo is more closely related to its state at $z=0$ rather than to properties of its progenitor in the past. 
These trends are reasonable in that the stellar mass and the gas metallicity are {\it integrated} properties over several billion years in the past, while the gas mass can be considered as an {\it instantaneous} property.

Readers should note that features with higher importances do not necessarily mean that these features are essential for the prediction. For example, the models highly rely on the subhalo maximum circular velocity and host halo virial radius when predicting a subhalo's stellar mass and gas metallicity. Although the models prefer to use these features as proxies for the dynamical mass of the subhalo and host halo, the models could achieve comparable performance without these parameters by utilizing other features with similar physical meanings (e.g., subhalo DM mass). Interestingly, \cite{2022MNRAS.513.5423M} show similar trends when predicting the stellar masses and stellar metallicities of galaxies out of DM halos. In their model, the feature importances of velocity dispersion and maximum circular velocity are more than five times higher than that of DM mass, with a similar peak observed at $z\sim1$. The pronounced preference for the virial radius of the host halo and the maximum circular velocity dispersion of subhalos, in comparison to other features related to dynamical mass, requires further investigation to elucidate the (possible) physical basis of this phenomenon. 

\begin{figure}
    \centering
    \includegraphics[width = 0.91\linewidth]{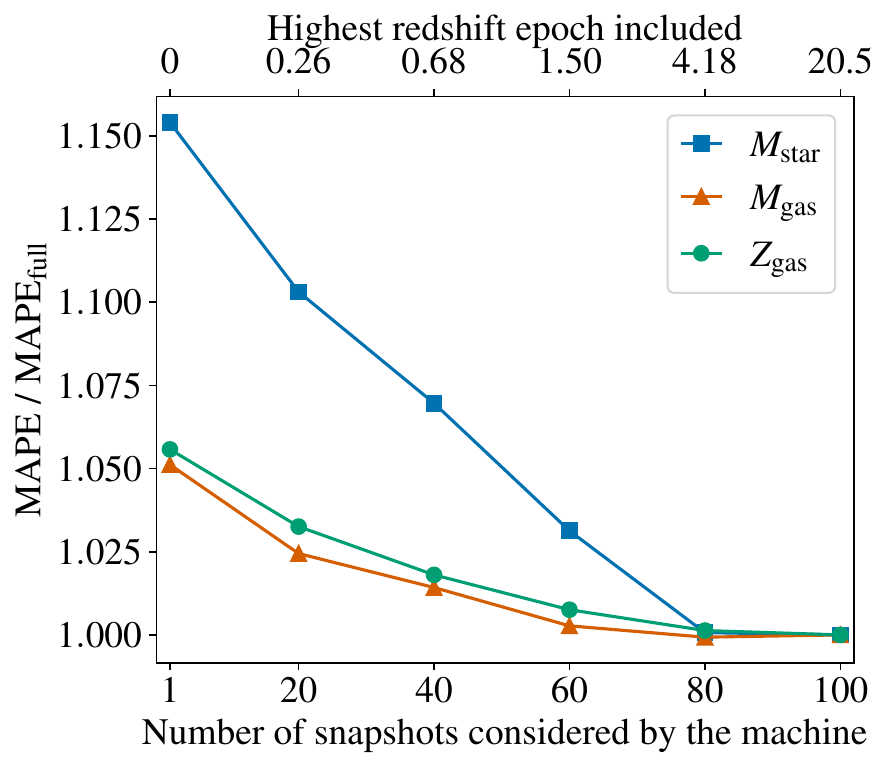}    
    \caption{The relative performance of the ML model that predicts the subhalo properties in a high-resolution simulation, as a function of the number of input snapshots fed to the model. 
    Training the model with only a single snapshot (i.e., only 23 features at $z=0$, instead of the full 2300 features) leads to 15\% worse prediction for stellar mass, as measured by the MAPE metric. 
    The model maintains its best performance as long as it considers more than 80 snapshots (i.e., all the snapshots below $z=4.18$). 
    As the number of input snapshots decreases, the prediction accuracy degrades significantly, especially for stellar mass. 
    See Section \ref{sec:feature_importance} for more details.}
    \label{fig:ml_ablation}
\end{figure}

To further verify the higher feature importances of high-z progenitors, we conduct an ablation study, as presented in Figure \ref{fig:ml_ablation}. 
We study the importance of input features from early epochs when predicting the stellar mass in a high-resolution simulation, by removing the high-$z$ features in the input of our ML models. 
For example, one can observe that training the model with only a single snapshot (i.e., only 23 features at $z=0$, instead of the full 2300 features at $20.5 \leq z \leq 0$) leads to performance degradation --- $\sim15\%$ worse for stellar mass and $\sim5\%$ worse for gas mass and gas metallicity, as measured by the MAPE metric. 
The result with the RMSE metric is almost identical with this figure. 
As we include more features from earlier epochs, the machine prediction becomes more accurate. 
The model reaches its best performance when it considers more than 80 snapshots out of the total 100 (i.e., all the snapshots below $z=4.18$). 
We conclude that the input features from the early epochs are important in our ML predictions, especially for the stellar mass prediction.  
This is reasonable because the subhalo's maximum circular velocity  (\texttt{SubhaloVmax}) or the host halo's virial radius (\texttt{Group\_R\_TopHat200}) in the early epoch plays a crucial role in this prediction, as illustrated in the left column of Figure \ref{fig:feature_importance}.

We carry out two more ablation studies:  {\it (i)} we train the ML model by excluding the local environmental information like the host halo's mass  (i.e., excluding the features marked with {\it ``(c)''}  in Table \ref{tab:features}), and, {\it (ii)} we train the ML model with only the features that can be obtained in DMO simulations (i.e., using only the features marked with {\it ``(b)''} in Table \ref{tab:features}).
The gas mass prediction is the most sensitive in both ablation studies; that is, $\sim 2.5 \%$ and $\sim13\%$ performance degradation for test {\it (i)} and {\it (ii)}, respectively. 
The inclusion of the baryonic properties is particularly essential for accurately predicting the gas mass of a subhalo in a high-resolution simulation. 
However, for the stellar mass prediction, the ML model maintains its performance even without the environmental features or the baryonic features --- only $\sim 1 \%$ and $\sim2.5\%$ performance degradation for test {\it (i)} and {\it (ii)}, respectively.  
We speculate that this is because the ML model can indirectly infer the subhalo's interactions with its surrounding environment by tracking the variations in subhalo features over time. 
For the stellar mass prediction, it seems more important to provide the information from the early epochs (as seen in Figure \ref{fig:ml_ablation}) than to provide the environmental features or the baryonic features.

\vspace{2mm}

\section{Summary and Conclusion}

To compare the subhalos between a high-resolution and low-resolution simulations, we have developed a unique subhalo matching algorithm which uses the main branches of the subhalo merger tree, instead of accessing the entire simulation snapshot data. 
Subhalos have been matched by pairing those with similar mass and position histories in the two simulations with differing resolutions (Section \ref{sec:matching}). \my{However, this approach shows a limitation in accuracy when pairing satellite subhalos.}
We have studied the resolution biases using two large cosmological simulations that offer resolution-dependent results, TNG100-1/-2 and TNG50-1/-2. 
In our {\it galaxy-by-galaxy} resolution study with our matching subhalo catalogs, the following resolution biases have been found in subhalo properties:
\begin{itemize}
\item The DM mass, velocity dispersion, and peculiar velocity of subhalos, which is largely determined by the gravitational interaction of DM, show strong resolution convergence in both TNG100 and TNG50 (Section \ref{sec:result_halo}; Figures \ref{fig:halo_feat} and \ref{fig:halo_feat-50}).
\item The stellar masses of subhalos in a high-resolution simulation (TNG100-1/TNG50-1) are $\lesssim 0.5$ dex higher than in their low-resolution counterparts (TNG100-2/TNG50-2), consistent with previous studies on TNG simulations.  
Since the subhalos' gas metallicities  are directly related to their stellar masses, subhalos in a high-resolution simulation have higher metallicities than their low-resolution counterparts (Section \ref{sec:result_halo}; panels {\it (a)} and {\it (b)} in Figures \ref{fig:mass_feat} and \ref{fig:mass_feat-50}).
\item The gas masses of subhalos with $M_{\rm gas} \sim 10^{8.5} \msun$ in a low-resolution simulation are higher than in their high-resolution counterparts, due to the lack of stars and stellar feedback in a low-resolution run, which would have otherwise expelled the gas out of the relatively shallow gravitational potential.  
The MBH masses of subhalos show relatively good resolution convergence, as they are directly linked to the halo masses (Section \ref{sec:result_halo}; panels {\it (c)} and {\it (d)} in Figures \ref{fig:mass_feat}  and \ref{fig:mass_feat-50}).
\end{itemize}

We have then investigated the resolution biases in the radial profiles of enclosed masses for central galaxies. 
Our findings in this {\it galaxy-by-galaxy} resolution study on radial profiles are as follows:
\begin{itemize}
\item The DM mass profiles show good resolution convergence, except in the region near the resolution limit (\my{$r \lesssim 4$ kpc}). 
The deficit in the stellar mass profiles in a low-resolution simulation (TNG100-2) compared to their counterparts in a high-resolution run (TNG100-1) is more pronounced, especially in \my{the inner regions} (Section \ref{sec:result_rc}; left two columns in Figure \ref{fig:rc_ratio}).  
\item The cold gas mass profiles show relatively good resolution convergence for galaxies with $M_{\rm DM} < 10^{12.5} \msun$, albeit with large galaxy-to-galaxy variance. However, for these galaxies, a deficit in hot gas mass is observed in the low-resolution simulation. In galaxies with $M_{\rm DM} \sim 10^{13}$, where AGN heating is significant, a deficit in cold gas mass is observed in the low-resolution simulation, while the hot gas mass shows good resolution convergence. 
The gas metallicity profiles also exhibit solid resolution convergence, except for the less massive central galaxies with $M_{\rm DM} \in [10^{10}, 10^{11}]\msun$  (Section \ref{sec:result_rc}; right two columns in Figure \ref{fig:rc_ratio}).
\end{itemize}

Finally,  using the matching catalog, we have developed an ML technique to construct a regression model that ``corrects'' subhalo properties in a low-resolution simulation, in order to mimic its counterpart's properties in a high-resolution simulation (Section \ref{sec:correcting}). 
This technique allows us to ``correct'' discrepancies arising from the differences in resolution, particularly for the stellar mass and gas mass of subhalos, \my{although its performance is lower in correcting gas metallicity.}   
The ML model has been applied to the subhalos in TNG300-1 which has the same resolution as TNG100-2, our low-resolution run in the training set. 
Our tests demonstrates that the galaxy scaling relations, such as the SMHM relation or the MZR from the low-resolution TNG300-1, can be ``corrected'' to mitigate resolution biases. 
Our ML models can be used for a variety of scientific purposes.
Large cosmological simulations like TNG300-1 are designed to sacrifice resolution in favor of a larger box size.  
However, we can still obtain a larger sample of galaxies with  improved accuracies and reduced resolution biases, with the help of our ML model.

An important aspect of our study is the versatility of the pipeline we have constructed. 
Both the subhalo matching algorithm and the correction model can be easily applied to any other large cosmological simulation. 
This makes it a valuable tool for testing {\it and} mitigating the resolution biases of different numerical codes and physics models. 
\my{The scripts used to perform the analysis and generate this manuscript, along with the matching catalogs and the corrected physical quantities of TNG300 subhalos, are available on GitHub\footnote{\url{https://github.com/JungMinyong/SubhaloMatching}} and archived in Zenodo \citep{jung_2024_10677470}.}

\vspace{2mm}
\begin{CJK*}{UTF8}{mj}

\acknowledgments
We thank Jun Yong Park, Yuan-Sen Ting, Myoungwon Jeon, Hyunmi Song, and Jihye Shin for their detailed comments and feedback on the early version of our manuscript. We would like to acknowledge the Korean Astronomy and Machine Learning (KAML) meeting series for providing a forum for fruitful discussions on the intersection of astronomy and machine learning. 
J. -h. K was supported by the National Research Foundation of Korea (NRF) grant funded by the Korea government (MSIT) (No. 2022M3K3A1093827 and No. 2023R1A2C1003244). His work was also supported by the National Institute of Supercomputing and Network/Korea Institute of Science and Technology Information with supercomputing resources including technical support, grants KSC-2020-CRE-0219, KSC-2021-CRE-0442 and KSC-2022-CRE-0355. S.E.H. was partly supported by the project 우주거대구조를 이용한 암흑우주 연구 (``Understanding Dark Universe Using Large Scale Structure of the Universe''), funded by the Ministry of Science. S.E.H. was also supported by the National Research Foundation of Korea(NRF) grant funded by the Korea government(MSIT) (No. 2022M3K3A1093827). J.L. was supported by the National Research Foundation of Korea (NRF-2021R1C1C2011626). J.K. was supported by a KIAS Individual grant (KG039603) via the Center for Advanced Computation at Korea Institute for Advanced Study. The IllustrisTNG simulations were undertaken with compute time awarded by the Gauss Centre for Supercomputing (GCS) under GCS Large-Scale Projects GCS-ILLU and GCS-DWAR on the GCS share of the supercomputer Hazel Hen at the High Performance Computing Center Stuttgart (HLRS), as well as on the machines of the Max Planck Computing and Data Facility (MPCDF) in Garching, Germany.

\end{CJK*}

\appendix
\section{Detailed Illustration of the Matching Process}\label{sec:matching_appendix}
\subsection{Training Step}\label{sec:training}

During the training process, our objective is to determine the parameters ($\gamma$, $w_{\rm mass}$, $w_{I}$, $b$, and $p$) in the similarity function (Eqs.(\ref{equation_train}) and (\ref{equation_train2})) from the TNG100-1 and TNG100-1-Dark simulations, where the subhalo matching catalogs between the two simulations are provided by the TNG Collaboration (Section \ref{sec:tng}). All subhalos in TNG100-1 and TNG100-1-Dark with $M_{\rm DM} > 2\times10^{9} \msun$ are used. We split these subhalos into two groups. The first group, the training set, consists of subhalos whose $z$-components of their positions are less than $0.8L_{\rm box}$. The remaining subhalos constitute the test set. The optimal parameters in Eq.(\ref{equation_train}) differ depending on the subhalo mass. Therefore, we use five distinct mass bins in units of $M_{\rm DM}/\msun$ given by [$2\times 10^9$, $5\times 10^9$), [$5\times 10^9$, $10^{10}$), [$10^{10}$, $3\times 10^{10}$), [$3\times 10^{10}$, $10^{11}$), [$10^{11}$, ${\infty}$). Each bin undergoes separate parameter training. 

For all subhalos in TNG100-1, we identify their counterparts in TNG100-1-Dark. This identification starts with subhalos in TNG100-1-Dark that have $M_{\rm DM} > 2\times10^{9} \msun$ and are located within a 5 Mpc/h sphere centered on the TNG100-1 subhalo's position. We then compute similarity scores, $S$, using Eq.(\ref{equation_train}) for these identified subhalos.\footnote{In this calculation, we remove the baryon contribution to the DM masses of subhalos in the DMO simulations} Lastly, we apply the softmax function ($\sigma_i(\bold{x}) = e^{x_i}/\sum_j e^{x_j}$) to determine the probabilities that the subhalos are the true counterparts. This probability subsequently helps us to derive the loss of the model with given parameters.

Training the model with the cross-entropy loss function ($-\sum_i y_i \log{\sigma_i (\bold{S})}$; where $y_i$ is the true label), we observed that $w_I$, the weight of a given snapshot $I$, fluctuates significantly between snapshots. We suspect these fluctuations stem from specific preferences for certain subhalos, rather than any inherent physical significance. Consequently, we have adopted several strategies to reduce model complexity and prevent overfitting, while still maintaining similar accuracy in model performance. First, we utilize the same weight $w_{{\rm snap}, I}$ for every four snapshots, training only 25 sets of weights and expanding these into 100 (for a total of the 100 snapshots). Second, we incorporate a regularization term into the cross-entropy loss function to penalize fluctuations between consecutive weights. This discourages the model from overly depending on selective information from specific time periods, while maintaining similar performance during the training phase. Without these two strategies, the training could potentially be skewed by a small number of subhalos that display specific tendencies due to their unique characteristics. The loss function is presented as follows:
\begin{equation}\label{eqn:loss}
    \mathcal{L} = - \sum_i y_i \log{\sigma_i (\bold{S})} + 0.01\sum_I \left|\log{{\frac{w_{I}}{w_{I+1}}}}\right|
\end{equation}
We train this model until the cross-entropy loss (without the regularization term in Eq.(\ref{eqn:loss})) starts to rise in the test set. 

\begin{figure*}[t]
    \centering
    \includegraphics[width=0.7\linewidth]{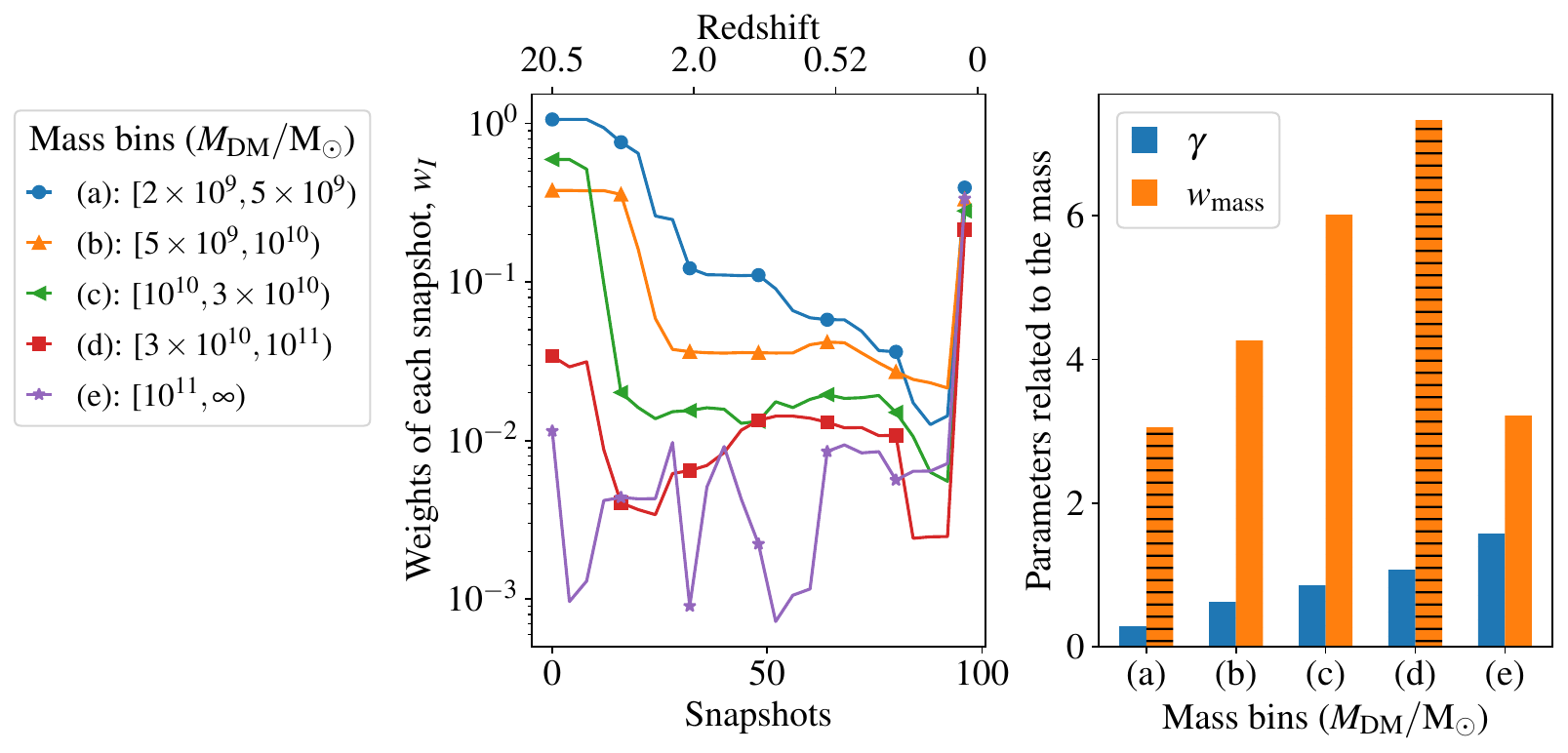}
    \caption{Parameters in the similarity function, Eqs.(\ref{equation_train}) and (\ref{equation_train2}). The left panel represents the weights of each snapshot, $w_I$. A higher value of $w_I$ indicates that the main progenitors in snapshot $I$ are more significant in the prediction. The right panel represents the parameters related to the mass differences, $\gamma$ and $w_{\rm mass}$. Higher values of $\gamma$ and $w_{\rm mass}$ suggest that mass differences between the two subhalos outweigh differences in their positions. We divide subhalos into five different mass bins and choose different parameters for each bin: (a) indicates the smallest mass bin ($2 \times 10^9 \leq M_{\rm DM}/{\rm M_{\odot}} < 5 \times 10^9$), and (e) indicates the most massive one ($2 \times 10^{11} \leq M_{\rm DM}/{\rm M_{\odot}}$). In the lower mass bins, the model prefers to give more weight to the progenitors at higher redshifts and prioritize position differences over mass differences. The higher mass bins show the opposite trend, giving more weight to the progenitors at low redshifts and emphasizing mass differences. See Section \ref{sec:training} for more details.}
    \label{fig:weights}
\end{figure*}

\vspace{1mm}

It is important to note that even if a subhalo follows the same growth history in the simulations of two different resolutions, the main progenitors in the higher resolution simulation could be identified at an earlier snapshot. This occurs because the halo finder can identify smaller subhalos in a simulation with higher resolution. Consequently, a ``true" pair could be incorrectly penalized because the corresponding progenitors in TNG100-2 have yet to be identified. Differences in the number of subhalos between TNG100-1 and TNG100-2 arise for those with $M_{\rm halo} \sim 10^9 \msun$. In order to minimize the impact of resolution differences between TNG100-1 and TNG100-2, we constrain our analysis to subhalos with masses greater than $2 \times 10^9 \msun$ (or 33 particles for TNG100-2). Additionally, we restrict their merger trees to only include the main progenitors with $M_{\rm DM} > 1.5 \times 10^{9} \msun$ (or 25 particles for TNG100-2). \my{These two restrictions are applied during both the training phase and the application phase.} 

\my{In the left panel of Figure \ref{fig:weights},} we display the weights of each snapshot across five mass bins, ranging from the lowest to the highest. Low-mass subhalos, which typically exhibit a relatively peaceful merger history and frequently coexist with many other subhalos of similar sizes and positions, tend to give more weight to information from the early epoch. In contrast, massive subhalos are more inclined to prioritize the position and mass information from the low redshifts over that from earlier times. The highest mass bin displays significant fluctuation due to a lack of subhalos. In the right panel of Figure \ref{fig:weights}, we show the two parameters: $\gamma$ and $w_{\rm mass}$, across the five mass bins. The model prefers lower $\gamma$ and $w_{\rm mass}$ values for the smaller mass bins, emphasizing position differences over mass differences. As the DM masses in low-mass subhalos could vary to some extent due to the baryonic physics, the model downplays the mass differences.  Consequently, the differences in parameters across the mass bins can be interpreted as the differences in growth histories of high-mass and low-mass galaxies.

We conduct the same training for the TNG50-1 and TNG50-1-Dark pair, and then apply it to the TNG50-1 and TNG50-2 pair. In this setup, we use subhalos with $M>2 \times 10^8 \msun$ (or 55 particles for TNG50-2), and only include the main progenitors with $M>1.5 \times 10^8 \msun$ (or 41 particles for TNG50-2). For the TNG50 suite, we use the following five mass bins in units of $M_{\rm DM}/\msun$ to be [$2\times 10^8, 5\times 10^8$), [$5\times 10^8, 10^9$), [$10^9, 10^{10}$), [$10^{10}, 10^{11}$), [$10^{11}$, ${\infty}$).

\subsection{Application Step}\label{sec:application}

The similarity function (Eq.(\ref{equation_train})) with the trained parameters is utilized to pair subhalos in TNG100-1 and TNG100-2. We apply the same method as described in Section \ref{sec:training} to match the subhalos in the two simulations. However, they are considered to be matched only if they are bijectively matched (i.e., both directions yield identical matching results). A step-by-step description of the application phase follows.
\begin{enumerate}
    \item For a target subhalo in TNG100-1, we identify all subhalos in TNG100-2 that have $M_{\rm DM} > 2 \times 10^9 \msun$ and are situated within a 5 Mpc/h sphere centered on the position of the subhalo in TNG100-1.
    \item We compute the similarity scores for the subhalos in TNG100-2. The probability associated with those subhalos is computed using the softmax function, $\sigma_i(\bold{S})$ for the $i$th subhalo. We then identify the subhalo with the highest score and save the corresponding probability.
    \item Steps 1-2 are repeated for all subhalos in TNG100-1. With this step, we construct a one-directional matching catalog (from TNG100-1 to TNG100-2).
    \item We repeat steps 1-3, but this time, starting from target subhalos in TNG100-2 and matching the corresponding subhalos in TNG100-1.
    \item We merge the two matching catalogs, retaining matches where both catalogs give identical results. We further narrow down the selections based on the following criteria: (i) The probability of the prediction is higher than \my{0.33} in both directions. (ii) The DM masses of the matched subhalo in both simulations exceed $3\times 10^9 \msun$ (or $3 \times 10^8\msun$ for TNG50).
\end{enumerate}

While this method ensures matched subhalos will have notably similar growth histories, it does not guarantee that the two halos indeed originated from the same DM region in the initial conditions. We provide further verification in Section \ref{sec:verifi}, demonstrating that nearly all subhalo pairs matched with this method indeed originate from similar dark matter regions in the initial conditions.  

   \begin{table*}
   \centering
   \begin{tabular}{lc|cccc}
       \toprule
       Target & Method & RMSE & MAPE & Bias & Pearson $\rho$ \\
       \midrule
       \multirow{3}{*}{\shortstack[l]{\textit{Stellar mass} \\(centrals)}} & Before correction (Figure \ref{fig:mass_feat}) & 0.654 & 0.069 & 0.4757& 0.919\\
        & After correction by gaussian process regression & 0.426 & 0.042 & $-0.0166$ & 0.927 \\
        & After correction by ML (Figure  \ref{fig:ml_gas}) & 0.203 & 0.020 & 0.0046 & 0.984\\
       \midrule
       \multirow{3}{*}{\shortstack[l]{\textit{Stellar mass} \\(satellites)}} & Before correction (Figure \ref{fig:mass_feat}) & 0.670 & 0.070 & 0.5329& 0.925\\
        & After correction by gaussian process regression & 0.399 & 0.037 & $0.0305$ & 0.928 \\
        & After correction by ML (Figure  \ref{fig:ml_gas}) & 0.204 & 0.019 & 0.0093 & 0.981\\
       \midrule
       \multirow{3}{*}{\shortstack[l]{\textit{Gas metallicity} \\(centrals)}} & Before correction (Figure \ref{fig:mass_feat}) & 0.521 & 0.195 & 0.3843 & 0.705\\
        & After correction by gaussian process regression & 0.269 & 0.087 & -0.0051 & 0.732\\
        & After correction by ML (Figure  \ref{fig:ml_gas}) & 0.178 & 0.059 & 0.0089 & 0.893\\
       \midrule
       \multirow{3}{*}{\shortstack[l]{\textit{Gas metallicity} \\(satellites)}} & Before correction (Figure \ref{fig:mass_feat}) & 0.448 & 0.174 & 0.3294 & 0.787\\
        & After correction by gaussian process regression & 0.228 & 0.081 & 0.0186 & 0.794\\
        & After correction by ML (Figure  \ref{fig:ml_gas}) & 0.163 & 0.058 & 0.0015 & 0.900\\
       \midrule
       \multirow{3}{*}{\shortstack[l]{\textit{Gas mass} \\(centrals)}} & Before correction (Figure \ref{fig:mass_feat}) & 0.473 & 0.043 & 0.2022 & 0.909\\
        & After correction by gaussian process regression & 0.352 & 0.031 & 0.0019 & 0.927\\
        & After correction by ML (Figure  \ref{fig:ml_gas}) & 0.260 & 0.021 & 0.0024 & 0.961\\
       \midrule
       \multirow{3}{*}{\shortstack[l]{\textit{Gas mass} \\(satellites)}} & Before correction (Figure \ref{fig:mass_feat}) & 0.516 & 0.044 & 0.0523 & 0.793\\
        & After correction by gaussian process regression & 0.465 & 0.040 & -0.0178 & 0.827\\
        & After correction by ML (Figure  \ref{fig:ml_gas}) & 0.344 & 0.028 & -0.0491 & 0.909\\
       \bottomrule
   \end{tabular}
   \vspace{1mm}   
   \caption{\my{Comparison of performance metrics for three subhalo properties: stellar mass, gas metallicity, and gas mass. These results are identical to Table \ref{tab:model_comparison}, but we separate the central and satellite subhalos. See Section \ref{sec:ml_sate} for more details.}}
   \label{tab:model_comparison_satellite}
   \vspace{1mm}
\end{table*}

\section{Machine learning correction of central and satellite subhalos}\label{sec:ml_sate}
\my{In Section \ref{sec:ml_result}, we present the correction of subhalo properties in the low-resolution simulation. Readers may expect that the satellite subhalos, which are less similar in terms of shared DM particles in each pair, show worse performance than the central subhalos. The performance metrics for central and satellite subhalos are listed in Table \ref{tab:model_comparison_satellite}. However, there is no systematic difference in the four metrics for stellar mass and gas metallicity. For the regression model predicting gas mass, the model performance is noticeably higher for the central subhalos, with all four metrics showing better performance. We conclude that the ML model can correct the satellite subhalos with consistent performance comparable to the central subhalos.}

\section{Subhalo Matching between Hydro and DMO simulations}\label{sec:dmo_comparison}
\begin{figure*}
    \centering
    \vspace{2mm}
    \includegraphics[width = 0.93\linewidth]{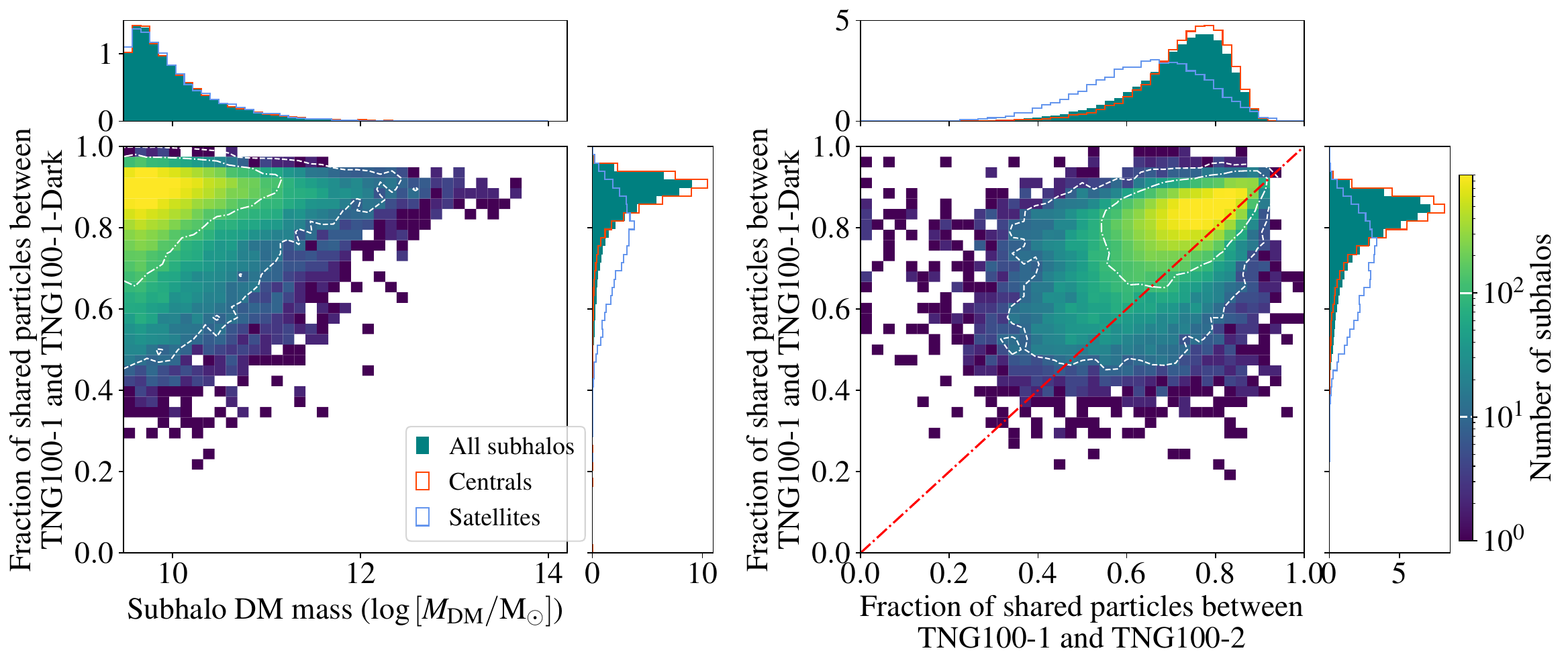}   
    \caption{\my{({\it left}) The fraction of DM particles in a TNG100-1-Dark (a DMO simulation) subhalo that have corresponding particles in its matching subhalo in TNG100-1. The marginal plots on the right and at the top display density histograms of the fractions and subhalo masses, respectively. In the {\it right} panel, the $y$-axis represents this fraction, while the $x$-axis corresponds to the fraction shared between TNG100-1 and TNG100-2. The matching pairs between the hydrodynamic and DMO simulations exhibit higher fractions than those between the high- and low-resolution simulations. See Section \ref{sec:dmo_comparison} for more details.}}   
    \vspace{1mm}
    \label{fig:predict_verif_dmo}
\end{figure*}

\my{We observe in Section \ref{sec:verifi} that the matching pairs we construct share a considerable amount of DM particles, even though we match subhalos by pairing those with similar growth histories. We now compare these results with those obtained from the subhalo pairs identified between DMO and hydrodynamic simulations, using the same subhalos randomly selected in Section \ref{sec:verifi}. In Figure \ref{fig:predict_verif_dmo}, we examine the DM fractions of subhalos in TNG100-1-Dark that have corresponding particles in their matching subhalos in TNG100-1, using the matching catalog provided by the TNG Collaboration. As this matching catalog directly utilizes the DM particle IDs, the fractions are mostly above 0.4. Moreover, these fractions are generally higher than those in comparisons between TNG100-1 and TNG100-2, even in pairs with fractions higher than 0.5.\footnote{\my{The fractions of particles in TNG100-1-Dark subhalos that are shared with TNG100-1 are {\it lower} than those in TNG100-1 subhalos shared with TNG100-1-Dark. This is because the DM mass of a subhalo in the DMO simulations is higher due to the absence of baryonic physics in the dwarf galaxy regime. The fractions are further slightly reduced when comparing subhalos in TNG100-2-Dark with those in TNG100-2, attributed to the decrease in resolution. Nonetheless, these fractions are still higher compared to the fraction obtained when comparing TNG100-1 and TNG100-2.}} Both central and satellite subhalos exhibit a similar drop in fractions between the high- and low-resolution simulations. The difference between the two distributions stems from intrinsic differences in the simulations, not from the matching method used. This suggests that the resolution difference has a greater impact on the similarity between the subhalo pairs than the presence of baryons.}

%

\vspace{5mm}





\bibliography{main}{}
\bibliographystyle{aasjournal}



\end{document}